\documentclass[a4paper,preprints,article,accept,oneauthor,pdftex]{Definitions/mdpi}
\firstpage{1}
\pubvolume{1}
\issuenum{1}
\articlenumber{0}
\pubyear{2021}
\copyrightyear{2021}
\datereceived{}
\dateaccepted{}
\datepublished{}
\hreflink{https://doi.org/} 
\pdfoutput=1

\usepackage{accents}

\newcommand{\lc}[1]{\accentset{\circ}{#1}}
\newcommand{\dd}{\mathrm{d}}
\newcommand{\DD}{\mathrm{D}}

\newcommand{\intprod}{\mathrel{\reflectbox{\rotatebox[origin=c]{180}{$\neg$}}}}

\Title{Variational Principles in Teleparallel Gravity Theories}

\TitleCitation{Variational principles in teleparallel gravity theories}

\Author{Manuel Hohmann $^{1}$*}

\AuthorNames{Manuel Hohmann}

\AuthorCitation{Hohmann, M}

\address{%
$^{1}$ \quad Laboratory of Theoretical Physics, Institute of Physics, University of Tartu, W. Ostwaldi 1, 50411 Tartu, Estonia; manuel.hohmann@ut.ee}

\corres{Correspondence: manuel.hohmann@ut.ee}

\abstract{We study the variational principle and derivation of the field equations for different classes of teleparallel gravity theories, using both their metric-affine and covariant tetrad formulations. These theories have in common that in addition to the tetrad or metric, they employ a flat connection as additional field variable, but differ by the presence of absence of torsion and nonmetricity for this independent connection. Besides the different underlying geometric formulation using a tetrad or metric as fundamental field variable, one has different choices to introduce the conditions of vanishing curvature, torsion and nonmetricity, either by imposing them a priori and correspondingly restricting the variation of the action when the field equations are derived, or by using Lagrange multipliers. Special care must be taken, since these conditions form non-holonomic constraints. Here we show explicitly that all of the aforementioned approaches are equivalent, and that the same set of field equations is obtained, independently of the choice of the geometric formulation and variation procedure. We further discuss consequences arising from the diffeomorphism invariance of the gravitational action, and show how they establish relations between the gravitational field equations.}

\keyword{teleparallel gravity; action principle; variation; Palatini formulation}

\begin{document}
\section{Introduction}\label{sec:intro}
Besides its most well-known description in terms of the curvature of the Levi-Civita connection, general relativity admits teleparallel descriptions, in which a flat connection is used to mediate the gravitational action in addition to the metric~\cite{BeltranJimenez:2019tjy}. While curvature is absent in these alternative approaches, the role of the gravitational field strength is attributed to the torsion~\cite{Einstein:1928,Aldrovandi:2013wha,Maluf:2013gaa} or the nonmetricity~\cite{Nester:1998mp,Adak:2005cd,Adak:2006rx,Mol:2014ooa,BeltranJimenez:2017tkd,Adak:2018vzk} of the flat connection, or to both~\cite{Jimenez:2019ghw,Bohmer:2021eoo}. Despite their different geometric structure, each of these different approaches leads to the same field equations for the metric degrees of freedom, so that they also share the same solutions for the metric, and are therefore regarded as equivalent. However, by considering modifications of these equivalent theories, in order to address open questions in gravity and cosmology, the equivalence is lost, and one finds new classes of gravity theories which cannot be obtained from modifying the curvature based description of general relativity~\cite{Cai:2015emx,Bahamonde:2017wwk,Hohmann:2018rwf,Hohmann:2018vle,Hohmann:2018dqh,Hohmann:2018ijr,Jarv:2018bgs,Runkla:2018xrv,Bahamonde:2019shr,Jimenez:2019ovq}.

Different geometric formulations are common for the aforementioned teleparallel gravity theories. One approach, which is commonly used in the general and symmetric teleparallel theories, in which nonmetricity is non-vanishing, is to consider the metric and a flat affine connection as fundamental field variables. This approach, which is set in the metric-affine geometry~\cite{Hehl:1994ue}, is also known as the Palatini approach~\cite{BeltranJimenez:2018vdo}. It has the advantage that the metric, which is physically observable through the motion of test matter, is used as a fundamental variable, and no additional gauge degrees of freedom are introduced. In contrast, in particular for metric teleparallel gravity theories, a description in terms of tetrads and a spin connection is more common~\cite{Krssak:2015oua,Golovnev:2017dox,Hohmann:2018rwf,Krssak:2018ywd}, which has the advantage that the derivation of the gravitational field equations from the action of a given theory is more straightforward, but it comes at the cost of introducing additional gauge degrees of freedom. In the case of metric teleparallel gravity, one finds that the spin connection is a pure gauge degree of freedom, and so it is also possible to a priori fix a specific gauge, called the Weitzenböck gauge, in which the spin connection vanishes, and to work with the tetrad only.

Different methods have been introduced in order to derive the gravitational field equations in the metric-affine and covariant tetrad formulations of teleparallel gravity theories. The necessity of these methods arises from the assumed flatness of the spin connection. Without taking this condition into account, and naively varying the action of a teleparallel gravity theory with respect to all connection components, hence treating them as independent fields, does not lead to the correct field equations~\cite{Golovnev:2017dox}. Two approaches are most common to address this problem. One possibility is to introduce Lagrange multipliers into the action, which impose the condition of vanishing curvature; the same method may be used to impose also the vanishing torsion or nonmetricity in the case of metric or symmetric teleparallel theories. While this approach is most straightforward, as it allows to treat all connection components as independent variables, with respect to which the action must be varied, it also turns out to be cumbersome, since the Lagrange multipliers constitute additional variables, which must be eliminated from the Euler-Lagrange equations, in order to obtain the field equations for the physical fields. This can be circumvented by a priori imposing the condition of vanishing curvature (and possibly also vanishing torsion or nonmetricity) on the connection, and allow only such variations which preserve this condition. This approach is essentially equivalent to integrating the flat connection, i.e., to assume that it is locally expressed as the derivative of a linear transformation, and to consider this linear transformation as the fundamental variable which is to be varied.

The multitude of geometric formulations and variation prescriptions in teleparallel gravity theories raises the question whether these approaches are equivalent, and give rise to the same field equations. This question is non-trivial, since the conditions of vanishing curvature and, depending on the choice of either the metric-affine or tetrad formulation, also of vanishing torsion or vanishing nonmetricity are non-holonomic, i.e., they depend not only on the dynamical fields, but also on their derivatives, and so care must be taken when the method of Lagrange multipliers is used~\cite{Ray:1966a,Ray:1966b,Flannery:2005}. In the case at hand, however, the constraints turn out to be semi-holonomic, as they exhibit an affine dependence on the field derivatives, and they turn out to be (locally) integrable. Hence, one may expect the different methods for deriving the field equations to be equivalent. This has been studied in the metric teleparallel case, where the equivalence between the restricted variation and Lagrange multiplier methods for deriving the field equations in the tetrad formulation has been discussed~\cite{Golovnev:2017dox}.

The aim of this article is to give an overview of the different formulations and variation prescriptions used for teleparallel gravity theories, and to show their equivalence for deriving the gravitational field equations. Here we follow a didactic approach, providing detailed steps for all necessary calculations. We show for each class of theories how the constraints of vanishing curvature, torsion and nonmetricity can be integrated, in order to obtain a restricted variation prescription. We then compare the result with the field equations obtained from the Lagrange multiplier method. For the latter we show that the Lagrange multipliers can be fully eliminated from the field equations, so that no unphysical dependence of the solution on the initial conditions for the Lagrange multipliers arises, which is a potential issue in the presence of non-holonomic constraints. We perform these calculations both in the metric-affine and tetrad formulations, and explicitly show their equivalence. For the latter, we make use of the language of differential forms, which turns out to be the most concise and compact.

The outline of this article is as follows. In section~\ref{sec:telegeo}, we give a brief overview of the different formulations used for teleparallel geometries, and define the notation which will be used throughout the article. The main part is given by section~\ref{sec:telegrav}, where we study the different classes of teleparallel gravity theories in the aforementioned formulations, and explicitly demonstrate the equivalence of these formulations for deriving the gravitational field equations. We conclude with a summary and outlook in section~\ref{sec:conclusion}. Throughout the article, we use the convention that Latin indices \(a, b, \ldots = 0, \ldots, 3\) denote Lorentz indices, while Greek indices \(\mu, \nu, \ldots = 0, \ldots, 3\) are spacetime indices, and the Minkowski metric is \(\eta_{ab} = \mathrm{diag}(-1, 1, 1, 1)\).

\section{Teleparallel geometries}\label{sec:telegeo}
We start our discussion of variational principles in teleparallel gravity with a brief review of the geometries and an exposition of the notational conventions we use in this article. Two different formulations are commonly used for teleparallel gravity theories. We discuss the tetrad formulation, which makes use of a tetrad and flat spin connection as fundamental fields, in section~\ref{ssec:tetradgeo}. Another approach, which we outline in section~\ref{ssec:metaffgeo}, is the metric-affine or Palatini formulation, in which a metric and a flat affine connection are used as fundamental field variables. The relation between these formulations is summarized in section~\ref{ssec:geomrel}.

\subsection{Tetrad formulation}\label{ssec:tetradgeo}
We start with a brief overview of the tetrad formulation of teleparallel geometry. The most concise way to describe this formulation makes use of the language of differential forms. In this language, the fundamental fields are given by the tetrad one-forms, \(\theta^a = \theta^a{}_{\mu}\dd x^{\mu}\) and the spin connection one-forms \(\omega^a{}_b = \omega^a{}_{b\mu}\dd x^{\mu}\). The tetrad is assumed to be invertible, with the inverse given by the vector fields \(e_a = e_a{}^{\mu}\partial_{\mu}\) satisfying
\begin{equation}
e_a \intprod \theta^b = e_a{}^{\mu}\theta^b{}_{\mu} = \delta_a^b\,.
\end{equation}
The spin connection is further restricted to be flat, i.e., has identically vanishing curvature
\begin{equation}\label{eq:spiconflat}
R^a{}_b = \dd\omega^a{}_b + \omega^a{}_c \wedge \omega^c{}_b \equiv 0\,.
\end{equation}
In the general teleparallel class of theories, no further restrictions are imposed on the tetrad and spin connection. This is different in the metric teleparallel setting, where one in addition assumes vanishing nonmetricity
\begin{equation}\label{eq:spiconnonmet}
Q_{ab} = \DD\eta_{ab} = \dd\eta_{ab} - \omega^c{}_a \wedge \eta_{cb} - \omega^c{}_b \wedge \eta_{ac}\,.
\end{equation}
Using \(\dd\eta_{ab} = 0\), vanishing nonmetricity implies that the spin connection is antisymmetric, \(\omega_{(ab)} = 0\), where we used the Minkowski metric to lower a Lorentz index. In contrast, in the symmetric teleparallel class of gravity theories, the nonmetricity may be non-vanishing, but one assumes vanishing torsion
\begin{equation}\label{eq:spicontors}
T^a = \DD\theta^a = \dd\theta^a + \omega^a{}_b \wedge \theta^b\,.
\end{equation}
Note that the curvature and torsion are two-forms, while the nonmetricity is a one-form.

It follows from the flatness condition~\eqref{eq:spiconflat} that the spin connection can locally (on a simply-connected region) be written in the form
\begin{equation}\label{eq:spiconinteg}
\omega^a{}_b = (\Omega^{-1})^a{}_c \wedge \dd\Omega^c{}_b\,,
\end{equation}
where the zero-forms \(\Omega^a{}_b\) form an invertible matrix. One finds that the curvature indeed vanishes, since
\begin{equation}
\dd\omega^a{}_b = \dd\left[(\Omega^{-1})^a{}_c \wedge \dd\Omega^c{}_b\right] = -(\Omega^{-1})^a{}_d \wedge \dd\Omega^d{}_c \wedge (\Omega^{-1})^c{}_e \wedge \dd\Omega^e{}_b = -\omega^a{}_c \wedge \omega^c{}_b\,.
\end{equation}
Another possibility to express the spin connection is through the decomposition
\begin{equation}\label{eq:spicondec}
\omega^a{}_b = \lc{\omega}^a{}_b + M^a{}_b = \lc{\omega}^a{}_b + K^a{}_b + L^a{}_b
\end{equation}
into the Levi-Civita spin connection
\begin{equation}\label{eq:levicivitaspi}
\lc{\omega}_{ab} = -\frac{1}{2}\left(e_b \intprod e_c \intprod \dd\theta_a + e_c \intprod e_a \intprod \dd\theta_b - e_a \intprod e_b \intprod \dd\theta_c\right) \wedge \theta^c\,,
\end{equation}
the contortion
\begin{equation}
K_{ab} = \frac{1}{2}\left(e_b \intprod e_c \intprod T_a + e_c \intprod e_a \intprod T_b - e_a \intprod e_b \intprod T_c\right) \wedge \theta^c
\end{equation}
and the disformation
\begin{equation}
L_{ab} = \frac{1}{2}\left(e_a \intprod Q_{bc} - e_b \intprod Q_{ac} - e_c \intprod Q_{ab}\right) \wedge \theta^c\,,
\end{equation}
where the latter two are subsumed in the distortion \(M^a{}_b\).

\subsection{Metric-affine formulation}\label{ssec:metaffgeo}
In the metric-affine or Palatini formulation of teleparallel gravity theories, for which we will use tensor component notation for convenience, the fundamental fields are a Lorentzian metric \(g_{\mu\nu}\) and an affine connection with coefficients \(\Gamma^{\mu}{}_{\nu\rho}\). As in the tetrad formulation, the connection is assumed to be flat,
\begin{equation}\label{eq:affconflat}
R^{\mu}{}_{\nu\rho\sigma} = \partial_{\rho}\Gamma^{\mu}{}_{\nu\sigma} - \partial_{\sigma}\Gamma^{\mu}{}_{\nu\rho} + \Gamma^{\mu}{}_{\tau\rho}\Gamma^{\tau}{}_{\nu\sigma} - \Gamma^{\mu}{}_{\tau\sigma}\Gamma^{\tau}{}_{\nu\rho} \equiv 0\,.
\end{equation}
Also in this formulation, this is the only condition which is imposed in the general teleparallel case. In the metric teleparallel case, one further imposes the condition of vanishing nonmetricity
\begin{equation}\label{eq:affconnonmet}
Q_{\mu\nu\rho} = \nabla_{\mu}g_{\nu\rho} = \partial_{\mu}g_{\nu\rho} - \Gamma^{\sigma}{}_{\nu\mu}g_{\sigma\rho} - \Gamma^{\sigma}{}_{\rho\mu}g_{\nu\sigma}\,,
\end{equation}
while in symmetric teleparallel gravity one demands that the torsion
\begin{equation}\label{eq:affcontors}
T^{\mu}{}_{\nu\rho} = \Gamma^{\mu}{}_{\rho\nu} - \Gamma^{\mu}{}_{\nu\rho}
\end{equation}
vanishes.

Similarly to the tetrad formulation, also the flatness~\eqref{eq:affconflat} can be integrated, in order to write the connection coefficients locally in the form
\begin{equation}\label{eq:affconinteg}
\Gamma^{\mu}{}_{\nu\rho} = (\Omega^{-1})^{\mu}{}_{\sigma}\partial_{\rho}\Omega^{\sigma}{}_{\nu}
\end{equation}
through a tensor field \(\Omega^{\mu}{}_{\nu}\). One easily checks that this satisfies
\begin{equation}
\begin{split}
\partial_{\rho}\Gamma^{\mu}{}_{\nu\sigma} - \partial_{\sigma}\Gamma^{\mu}{}_{\nu\rho} &= \partial_{\rho}\left[(\Omega^{-1})^{\mu}{}_{\tau}\partial_{\sigma}\Omega^{\tau}{}_{\nu}\right] - \partial_{\sigma}\left[(\Omega^{-1})^{\mu}{}_{\tau}\partial_{\rho}\Omega^{\tau}{}_{\nu}\right]\\
&= \partial_{\rho}(\Omega^{-1})^{\mu}{}_{\tau}\partial_{\sigma}\Omega^{\tau}{}_{\nu} + (\Omega^{-1})^{\mu}{}_{\tau}\partial_{\rho}\partial_{\sigma}\Omega^{\tau}{}_{\nu} - \partial_{\sigma}(\Omega^{-1})^{\mu}{}_{\tau}\partial_{\rho}\Omega^{\tau}{}_{\nu} - (\Omega^{-1})^{\mu}{}_{\tau}\partial_{\sigma}\partial_{\rho}\Omega^{\tau}{}_{\nu}\\
&= -(\Omega^{-1})^{\mu}{}_{\lambda}\partial_{\rho}\Omega^{\lambda}{}_{\omega}(\Omega^{-1})^{\omega}{}_{\tau}\partial_{\sigma}\Omega^{\tau}{}_{\nu} + (\Omega^{-1})^{\mu}{}_{\lambda}\partial_{\sigma}\Omega^{\lambda}{}_{\omega}(\Omega^{-1})^{\omega}{}_{\tau}\partial_{\rho}\Omega^{\tau}{}_{\nu}\\
&= -\Gamma^{\mu}{}_{\tau\rho}\Gamma^{\tau}{}_{\nu\sigma} + \Gamma^{\mu}{}_{\tau\sigma}\Gamma^{\tau}{}_{\nu\rho}\,,
\end{split}
\end{equation}
so that the resulting connection is indeed flat. Finally, one may also decompose the coefficients of the affine connection in the form
\begin{equation}\label{eq:affcondec}
\Gamma^{\rho}{}_{\mu\nu} = \lc{\Gamma}^{\rho}{}_{\mu\nu} + M^{\rho}{}_{\mu\nu} = \lc{\Gamma}^{\rho}{}_{\mu\nu} + K^{\rho}{}_{\mu\nu} + L^{\rho}{}_{\mu\nu}\,,
\end{equation}
where in addition to the coefficients
\begin{equation}\label{eq:levicivitaaff}
\lc{\Gamma}^{\mu}{}_{\nu\rho} = \frac{1}{2}g^{\mu\sigma}\left(\partial_{\nu}g_{\sigma\rho} + \partial_{\rho}g_{\nu\sigma} - \partial_{\sigma}g_{\nu\rho}\right)
\end{equation}
of the Levi-Civita connection, we have the contortion tensor
\begin{equation}\label{eq:contor}
K^{\mu}{}_{\nu\rho} = \frac{1}{2}\left(T_{\nu}{}^{\mu}{}_{\rho} + T_{\rho}{}^{\mu}{}_{\nu} - T^{\mu}{}_{\nu\rho}\right)\,,
\end{equation}
as well as the disformation tensor
\begin{equation}\label{eq:disfor}
L^{\mu}{}_{\nu\rho} = \frac{1}{2}\left(Q^{\mu}{}_{\nu\rho} - Q_{\nu}{}^{\mu}{}_{\rho} - Q_{\rho}{}^{\mu}{}_{\nu}\right)\,,
\end{equation}
whose sum is the distortion \(M^{\mu}{}_{\nu\rho}\).

Finally, it is helpful to note the first Bianchi identity, which reads
\begin{equation}\label{eq:bianchioneflat}
\nabla_{[\sigma}T^{\rho}{}_{\mu\nu]} + T^{\rho}{}_{\tau[\sigma}T^{\tau}{}_{\mu\nu]} = 0
\end{equation}
in the absence of curvature. From this relation one obtains the contracted Bianchi identity
\begin{equation}\label{eq:bianchioneflatc}
3\nabla_{[\tau}T^{\tau}{}_{\rho\sigma]} = -3T^{\tau}{}_{\omega[\tau}T^{\omega}{}_{\rho\sigma]} = T^{\tau}{}_{\tau\omega}T^{\omega}{}_{\rho\sigma} + T^{\tau}{}_{\rho\omega}T^{\omega}{}_{\sigma\tau} - T^{\tau}{}_{\sigma\omega}T^{\omega}{}_{\rho\tau} = T^{\tau}{}_{\tau\omega}T^{\omega}{}_{\rho\sigma}\,,
\end{equation}
where the last two terms cancel each other due to symmetry. We will make use of this geometric identity in later calculations.

\subsection{Relation between different formulations}\label{ssec:geomrel}
We finally summarize how the two different formulations of teleparallel geometries shown above are related. For this purpose, we express the metric
\begin{equation}\label{eq:metric}
g_{\mu\nu} = \eta_{ab}\theta^a{}_{\mu}\theta^b{}_{\nu}
\end{equation}
and the coefficients
\begin{equation}\label{eq:affcon}
\Gamma^{\mu}{}_{\nu\rho} = e_a{}^{\mu}(\partial_{\rho}\theta^a{}_{\nu} + \omega^a{}_{b\rho}\theta^b{}_{\nu})
\end{equation}
of the affine connection in terms of the tetrad and the spin connection. While the metric and affine connection are uniquely determined from the tetrad and spin connection, the converse is not true: one obtains the same metric and affine connection also from the tetrad and spin connection
\begin{equation}\label{eq:finloclortrans}
\theta'^a = \Lambda^a{}_b\theta^b\,, \quad
\omega'^a{}_b = \Lambda^a{}_c(\Lambda^{-1})^d{}_b\omega^c{}_d + \Lambda^a{}_c\dd(\Lambda^{-1})^c{}_b\,,
\end{equation}
which are related to the original field variables by a local Lorentz transformation \(\Lambda^a{}_b\) satisfying
\begin{equation}
\eta_{ab}\Lambda^a{}_c\Lambda^b{}_d = \eta_{cd}\,.
\end{equation}
This Lorentz gauge invariance of the tetrad formulation must be taken into account when the field equations are derived from the gravitational action, which will be done in the following section.

\section{Teleparallel gravity actions and field equations}\label{sec:telegrav}
Using the mathematical foundations laid out in the previous section, we now study the variation of teleparallel gravity actions and the derivation of the corresponding field equations in the different geometric formulations. In section~\ref{ssec:metaffgrav}, we use the metric-affine formulation, and we perform all calculations using the language of tensor components. For the tetrad formulation discussed in section~\ref{ssec:tetradgrav}, it turns out to be more convenient to work with differential forms instead. We relate both formulations to each other and show their equivalence in section~\ref{ssec:gravrel}.

\subsection{Metric-affine formulation}\label{ssec:metaffgrav}
We start our derivation in the metric-affine formulation. Before we discuss specific classes of teleparallel gravity theories, which are defined by the presence or absence of torsion and nonmetricity, we give a number of general remarks and definitions in section~\ref{sssec:metaffvar}. These are used in the following sections. In particular, in section~\ref{sssec:metaffgen} we study general teleparallel gravity, continue with metric teleparallel gravity in section~\ref{sssec:metaffmet}, and conclude with symmetric teleparallel gravity in section~\ref{sssec:metaffsym}.

\subsubsection{General action and variation}\label{sssec:metaffvar}
We first consider the metric-affine formulation. Here we split the teleparallel gravity action in the general form
\begin{equation}\label{eq:metaffactsplit}
S = S_{\text{g}}[g, \Gamma] + S_{\text{m}}[g, \psi] + S_{\text{l}}[g, \Gamma, r, t, q]\,,
\end{equation}
into a gravitational part \(S_{\text{g}}\), a matter part \(S_{\text{m}}\) and an optional Lagrange multiplier part \(S_{\text{l}}\). The gravitational part of the action depends on the metric and the affine connection, so that its variation takes the general form
\begin{equation}\label{eq:metricgravactvar}
\delta S_{\text{g}} = -\int_M\left(\frac{1}{2}W^{\mu\nu}\delta g_{\mu\nu} + Y_{\mu}{}^{\nu\rho}\delta\Gamma^{\mu}{}_{\nu\rho}\right)\sqrt{-g}\dd^4x\,,
\end{equation}
after eliminating any derivatives acting on the variations \(\delta g_{\mu\nu}\) and \(\delta\Gamma^{\mu}{}_{\nu\rho}\) using integration by parts. The tensorial quantities \(W^{\mu\nu}\) and \(Y_{\mu}{}^{\nu\rho}\) will be the central ingredient to the derivation of the gravitational field equations. The gravitational part of the action is complemented by a matter part, for which we assume that it depends only on the metric \(g_{\mu\nu}\) and a set of matter fields \(\psi^I\), which can be any tensor fields, but not on the independent affine connection defined by \(\Gamma^{\mu}{}_{\nu\rho}\). It follows that the variation of the matter action can be written in the form
\begin{equation}\label{eq:metricmatactvar}
\delta S_{\text{m}} = \int_M\left(\frac{1}{2}\Theta^{\mu\nu}\delta g_{\mu\nu} + \Psi_I\delta\psi^I\right)\sqrt{-g}\dd^4x\,,
\end{equation}
where \(\Theta_{\mu\nu}\) is the familiar Hilbert energy-momentum tensor, while \(\Psi_I = 0\) are the (tensorial) matter field equations. Note that one could also relax this assumption and allow for a direct coupling of matter to the teleparallel connection, by including a corresponding term \(H_{\mu}{}^{\nu\rho}\delta\Gamma^{\mu}{}_{\nu\rho}\) in the matter action~\cite{Harko:2018gxr,Lobo:2019xwp}. We do not consider this term here, but it can easily incorporated into the calculations shown here by complementing \(Y_{\mu}{}^{\nu\rho}\) in the field equations with \(H_{\mu}{}^{\nu\rho}\), in the same way as \(\Theta^{\mu\nu}\) complements \(W^{\mu\nu}\). The precise steps of deriving the field equations depend on the choice of the underlying geometric framework and the method of variation, and will be detailed below, where we also specify the Lagrange multiplier term \(S_{\text{l}}\) and its variation.

As another restriction, we assume that each of the aforementioned parts of the action is separately invariant under diffeomorphisms. Note that under an infinitesimal diffeomorphism generated by a vector field \(X^{\mu}\), the metric and connection coefficients change as~\cite{Yano:1957lda}
\begin{equation}
\delta g_{\mu\nu} = (\mathcal{L}_Xg)_{\mu\nu} = 2\lc{\nabla}_{(\mu}X_{\nu)}\,, \quad
\delta\Gamma^{\mu}{}_{\nu\rho} = (\mathcal{L}_X\Gamma)^{\mu}{}_{\nu\rho} = \nabla_{\rho}(\nabla_{\nu}X^{\mu} - T^{\mu}{}_{\nu\sigma}X^{\sigma})
\end{equation}
in the absence of curvature. Together with the Lie derivative of the metric fields \(\psi^I\), the induced variation of the matter action is thus given by
\begin{equation}
0 = \delta_XS_{\text{m}} = \int_M\left(\Theta^{\mu\nu}\lc{\nabla}_{\mu}X_{\nu} + \Psi_I\mathcal{L}_X\psi^I\right)\sqrt{-g}\dd^4x = \int_M\left(-\lc{\nabla}_{\mu}\Theta^{\mu\nu}X_{\nu} + \Psi_I\mathcal{L}_X\psi^I\right)\sqrt{-g}\dd^4x\,,
\end{equation}
where we used the symmetry of the energy-momentum tensor to omit the symmetrization brackets on the lower indices. On-shell, i.e., when the matter field equations \(\Psi_I = 0\) are satisfied, this implies the energy-momentum conservation
\begin{equation}\label{eq:metaffdiffinvmat}
\lc{\nabla}_{\mu}\Theta^{\mu\nu} = 0\,.
\end{equation}
For the gravitational part of the action, the corresponding variation reads
\begin{equation}
\begin{split}
0 = \delta_XS_{\text{g}} &= -\int_M\left[W^{\mu\nu}\lc{\nabla}_{\mu}X_{\nu} + Y_{\mu}{}^{\nu\rho}\nabla_{\rho}(\nabla_{\nu}X^{\mu} - T^{\mu}{}_{\nu\sigma}X^{\sigma})\right]\sqrt{-g}\dd^4x\\
&= \int_M\left[\lc{\nabla}_{\nu}\left(W_{\mu}{}^{\nu} - \nabla_{\rho}Y_{\mu}{}^{\nu\rho} + Y_{\mu}{}^{\nu\rho}M^{\sigma}{}_{\rho\sigma}\right) + \left(\nabla_{\sigma}Y_{\nu}{}^{\rho\sigma} - Y_{\nu}{}^{\rho\sigma}M^{\tau}{}_{\sigma\tau}\right)M^{\nu}{}_{\rho\mu}\right]\sqrt{-g}\dd^4x\,,
\end{split}
\end{equation}
after integration by parts and introducing distortion tensors in order to simplify the resulting covariant derivatives. One thus obtains the equation
\begin{equation}\label{eq:metaffdiffinvgrav}
\lc{\nabla}_{\nu}\left(W_{\mu}{}^{\nu} - \nabla_{\rho}Y_{\mu}{}^{\nu\rho} + Y_{\mu}{}^{\nu\rho}M^{\sigma}{}_{\rho\sigma}\right) + \left(\nabla_{\sigma}Y_{\nu}{}^{\rho\sigma} - Y_{\nu}{}^{\rho\sigma}M^{\tau}{}_{\sigma\tau}\right)M^{\nu}{}_{\rho\mu} = 0\,,
\end{equation}
which is a geometric identity for any diffeomorphism invariant teleparallel gravity action, i.e., it is satisfied for any metric and flat, affine connection, independently of whether these satisfy the field equations of the theory or not.

\subsubsection{General teleparallel gravity}\label{sssec:metaffgen}
We now come to the derivation of the field equations in the general teleparallel class of gravity theories, where the affine connection is assumed to be flat by the condition~\eqref{eq:affconflat}, from the action~\eqref{eq:metaffactsplit}. Note that simply taking the variation~\eqref{eq:metricgravactvar} of the gravitational action with respect to the connection coefficients would not result in the correct field equations, since it does not take into account the flatness condition. This condition can be imposed on the variation in different ways. One possibility is to restrict the variation such that the flatness~\eqref{eq:affconflat} is preserved; another approach makes use of Lagrange multipliers instead. In the following, we display both approaches, and show that they lead to the same field equations.

We start with the approach of restricted variation. For this purpose, note that the variation of the curvature tensor of the affine connection can be written in the simple form
\begin{equation}\label{eq:metaffvarflat}
\delta R^{\mu}{}_{\nu\rho\sigma} = \nabla_{\rho}\delta\Gamma^{\mu}{}_{\nu\sigma} - \nabla_{\sigma}\delta\Gamma^{\mu}{}_{\nu\rho} + T^{\tau}{}_{\rho\sigma}\delta\Gamma^{\mu}{}_{\nu\tau}
\end{equation}
in terms of the covariant derivative of the variation of the connection coefficients, which is possible since the latter form the components of a tensor field. In order to satisfy this constraint, one considers a restricted variation of the form
\begin{equation}\label{eq:metaffflatvar}
\delta\Gamma^{\mu}{}_{\nu\rho} = \nabla_{\rho}\xi^{\mu}{}_{\nu}\,,
\end{equation}
which can be obtained from the integral form~\eqref{eq:affconinteg} by setting
\begin{equation}
\delta\Omega^{\mu}{}_{\nu} = \Omega^{\mu}{}_{\rho}\xi^{\rho}{}_{\nu}\,.
\end{equation}
Inserting this ansatz into the flatness condition~\eqref{eq:metaffvarflat}, one finds that it indeed yields
\begin{equation}
\delta R^{\mu}{}_{\nu\rho\sigma} = \nabla_{\rho}\nabla_{\sigma}\xi^{\mu}{}_{\nu} - \nabla_{\sigma}\nabla_{\rho}\xi^{\mu}{}_{\nu} + T^{\tau}{}_{\rho\sigma}\nabla_{\tau}\xi^{\mu}{}_{\nu} = R^{\mu}{}_{\tau\rho\sigma}\xi^{\tau}{}_{\nu} - R^{\tau}{}_{\nu\rho\sigma}\xi^{\mu}{}_{\tau} = 0\,,
\end{equation}
using the well-known expression for the commutator of covariant derivatives in terms of the curvature and torsion. It follows that the variation of the total action with respect to the spin connection reads
\begin{equation}
\begin{split}
\delta_{\Gamma}S &= -\int_MY_{\mu}{}^{\nu\rho}\nabla_{\rho}\xi^{\mu}{}_{\nu}\sqrt{-g}\dd^4x\\
&= \int_M\left(\lc{\nabla}_{\rho}Y_{\mu}{}^{\nu\rho} - Y_{\tau}{}^{\nu\rho}M^{\tau}{}_{\mu\rho} + Y_{\mu}{}^{\tau\rho}M^{\nu}{}_{\tau\rho}\right)\xi^{\mu}{}_{\nu}\sqrt{-g}\dd^4x\\
&= \int_M\left(\nabla_{\rho}Y_{\mu}{}^{\nu\rho} - Y_{\mu}{}^{\nu\tau}M^{\rho}{}_{\tau\rho}\right)\xi^{\mu}{}_{\nu}\sqrt{-g}\dd^4x\,,
\end{split}
\end{equation}
using integration by parts with the Levi-Civita covariant derivative, and so one obtains the connection field equation
\begin{equation}\label{eq:affcongenfield}
\nabla_{\rho}Y_{\mu}{}^{\nu\rho} - Y_{\mu}{}^{\nu\tau}M^{\rho}{}_{\tau\rho} = 0\,,
\end{equation}
where the trace of the distortion tensor is given by
\begin{equation}
M^{\rho}{}_{\tau\rho} = T^{\rho}{}_{\rho\tau} - \frac{1}{2}Q_{\tau\rho}{}^{\rho}\,.
\end{equation}
With the help of this relation, it is possible to rewrite the connection field equation by introducing the tensor density
\begin{equation}\label{eq:affconvardens}
\tilde{Y}_{\mu}{}^{\nu\rho} = Y_{\mu}{}^{\nu\rho}\sqrt{-g}\,.
\end{equation}
Using the covariant derivative
\begin{equation}
\nabla_{\mu}\sqrt{-g} = \frac{1}{2}Q_{\mu\nu}{}^{\nu}\sqrt{-g}\,,
\end{equation}
one has
\begin{equation}
\nabla_{\rho}\tilde{Y}_{\mu}{}^{\nu\rho} - \tilde{Y}_{\mu}{}^{\nu\tau}T^{\rho}{}_{\rho\tau} = 0\,,
\end{equation}
since the nonmetricity cancels~\cite{BeltranJimenez:2018vdo,Jimenez:2019ghw}. Further, one calculates the variation of the action with respect to the metric, which reads
\begin{equation}
\delta_gS = \frac{1}{2}\int_M(\Theta^{\mu\nu} - W^{\mu\nu})\delta g_{\mu\nu}\sqrt{-g}\dd^4x\,,
\end{equation}
so that one finds the corresponding field equation
\begin{equation}\label{eq:metricgenfield}
W_{\mu\nu} = \Theta_{\mu\nu}\,.
\end{equation}
We remark that this field equation is symmetric in its two indices, so that it has 10 independent components, whereas the connection field equations have no particular symmetry, and so have 16 independent components, which results in a total of 26 equations.

Instead of a priori imposing that the curvature, and hence its variation vanishes, one may also implement this constraint on the connection by adding a Lagrange multiplier term
\begin{equation}
S_{\text{l}} = \int_Mr_{\mu}{}^{\nu\rho\sigma}R^{\mu}{}_{\nu\rho\sigma}\sqrt{-g}\dd^4x
\end{equation}
to the action. Variation with respect to the Lagrange multiplier \(r_{\mu}{}^{\nu\rho\sigma}\) then yields the flatness condition~\eqref{eq:affconflat}, while the variation with respect to the metric is unaltered, and so yields the same equation~\eqref{eq:metricgenfield} as before. In this approach, the connection field equation is obtained by varying the action with respect to all components \(\Gamma^{\mu}{}_{\nu\rho}\) of the connection coefficients, which yields
\begin{equation}
\begin{split}
\delta_{\Gamma}S &= \int_M\left[r_{\mu}{}^{\nu\rho\sigma}\left(\nabla_{\rho}\delta\Gamma^{\mu}{}_{\nu\sigma} - \nabla_{\sigma}\delta\Gamma^{\mu}{}_{\nu\rho} + T^{\tau}{}_{\rho\sigma}\delta\Gamma^{\mu}{}_{\nu\tau}\right) - Y_{\mu}{}^{\nu\rho}\delta\Gamma^{\mu}{}_{\nu\rho}\right]\sqrt{-g}\dd^4x\\
&= \int_M\left(2\nabla_{\sigma}r_{\mu}{}^{\nu[\rho\sigma]} - 2r_{\mu}{}^{\nu[\rho\sigma]}M^{\tau}{}_{\sigma\tau} - 2r_{\mu}{}^{\nu[\sigma\tau]}M^{\rho}{}_{\sigma\tau} - Y_{\mu}{}^{\nu\rho}\right)\delta\Gamma^{\mu}{}_{\nu\rho}\sqrt{-g}\dd^4x\,,
\end{split}
\end{equation}
where we have once again used integration by parts, as well as the relation
\begin{equation}
T^{\mu}{}_{\nu\rho} = M^{\mu}{}_{\rho\nu} - M^{\mu}{}_{\nu\rho}\,.
\end{equation}
The field equations therefore read
\begin{equation}\label{eq:lagmulgenfield}
2\nabla_{\sigma}r_{\mu}{}^{\nu[\rho\sigma]} - 2r_{\mu}{}^{\nu[\rho\sigma]}M^{\tau}{}_{\sigma\tau} - 2r_{\mu}{}^{\nu[\sigma\tau]}M^{\rho}{}_{\sigma\tau} - Y_{\mu}{}^{\nu\rho} = 0\,.
\end{equation}
In order to eliminate the Lagrange multiplier, one may consider the covariant divergence of this equation. The first term then yields the commutator of two covariant derivatives, and so becomes
\begin{equation}
2\nabla_{\rho}\nabla_{\sigma}r_{\mu}{}^{\nu[\rho\sigma]} = 2\nabla_{[\rho}\nabla_{\sigma]}r_{\mu}{}^{\nu\rho\sigma} = -T^{\tau}{}_{\rho\sigma}\nabla_{\tau}r_{\mu}{}^{\nu\rho\sigma} = 2M^{\rho}{}_{\sigma\tau}\nabla_{\rho}r_{\mu}{}^{\nu[\sigma\tau]}\,.
\end{equation}
One therefore arrives at the equation
\begin{equation}
\begin{split}
0 &= \nabla_{\rho}\left(Y_{\mu}{}^{\nu\rho} + 2r_{\mu}{}^{\nu[\rho\sigma]}M^{\tau}{}_{\sigma\tau} + 2r_{\mu}{}^{\nu[\sigma\tau]}M^{\rho}{}_{\sigma\tau}\right) - 2M^{\rho}{}_{\sigma\tau}\nabla_{\rho}r_{\mu}{}^{\nu[\sigma\tau]}\\
&= \nabla_{\rho}Y_{\mu}{}^{\nu\rho} + \nabla_{\rho}r_{\mu}{}^{\nu[\rho\sigma]}(2T^{\tau}{}_{\tau\sigma} - Q_{\sigma\tau}{}^{\tau}) + r_{\mu}{}^{\nu[\rho\sigma]}\nabla_{\rho}(2T^{\tau}{}_{\tau\sigma} - Q_{\sigma\tau}{}^{\tau}) - r_{\mu}{}^{\nu[\rho\sigma]}\nabla_{\tau}T^{\tau}{}_{\rho\sigma}\\
&= \nabla_{\rho}Y_{\mu}{}^{\nu\rho} + \nabla_{\rho}r_{\mu}{}^{\nu[\rho\sigma]}(2T^{\tau}{}_{\tau\sigma} - Q_{\sigma\tau}{}^{\tau}) - r_{\mu}{}^{\nu[\rho\sigma]}(3\nabla_{[\tau}T^{\tau}{}_{\rho\sigma]} + \nabla_{[\rho}Q_{\sigma]\tau}{}^{\tau})\,.
\end{split}
\end{equation}
Two substitutions can be applied to the last term. We replace the covariant derivative of the torsion tensor by using the contracted Bianchi identity~\eqref{eq:bianchioneflatc}. Further, the antisymmetric part of the covariant derivative of the nonmetricity reads
\begin{equation}\label{eq:nonmetasymder}
\nabla_{[\rho}Q_{\sigma]\tau}{}^{\tau} = \nabla_{[\rho}(g^{\omega\tau}\nabla_{\sigma]}g_{\omega\tau}) = -\frac{1}{2}T^{\mu}{}_{\rho\sigma}g^{\omega\tau}\nabla_{\mu}g_{\omega\tau} = -\frac{1}{2}T^{\mu}{}_{\rho\sigma}Q_{\mu\tau}{}^{\tau}\,.
\end{equation}
With these substitutions, the field equation becomes
\begin{equation}
\begin{split}
0 &= \nabla_{\rho}Y_{\mu}{}^{\nu\rho} + \nabla_{\rho}r_{\mu}{}^{\nu[\rho\sigma]}(2T^{\tau}{}_{\tau\sigma} - Q_{\sigma\tau}{}^{\tau}) - r_{\mu}{}^{\nu[\rho\sigma]}T^{\omega}{}_{\rho\sigma}\left(T^{\tau}{}_{\tau\omega} - \frac{1}{2}Q_{\omega\tau}{}^{\tau}\right)\\
&= \nabla_{\rho}Y_{\mu}{}^{\nu\rho} - 2\nabla_{\sigma}r_{\mu}{}^{\nu[\rho\sigma]}M^{\tau}{}_{\rho\tau} + 2r_{\mu}{}^{\nu[\rho\sigma]}M^{\omega}{}_{\rho\sigma}M^{\tau}{}_{\omega\tau}\,.
\end{split}
\end{equation}
To finally eliminate the Lagrange multipliers, we contract the original equation~\eqref{eq:lagmulgenfield} with the trace \(M^{\tau}{}_{\rho\tau}\) of the distortion, to obtain
\begin{equation}
\begin{split}
0 &= 2\nabla_{\sigma}r_{\mu}{}^{\nu[\rho\sigma]}M^{\tau}{}_{\rho\tau} - 2r_{\mu}{}^{\nu[\rho\sigma]}M^{\tau}{}_{\sigma\tau}M^{\omega}{}_{\rho\omega} - 2r_{\mu}{}^{\nu[\sigma\tau]}M^{\rho}{}_{\sigma\tau}M^{\omega}{}_{\rho\omega} - M^{\tau}{}_{\rho\tau}Y_{\mu}{}^{\nu\rho}\\
&= 2\nabla_{\sigma}r_{\mu}{}^{\nu[\rho\sigma]}M^{\tau}{}_{\rho\tau} - 2r_{\mu}{}^{\nu[\rho\sigma]}M^{\omega}{}_{\rho\sigma}M^{\tau}{}_{\omega\tau} - M^{\tau}{}_{\rho\tau}Y_{\mu}{}^{\nu\rho}\,,
\end{split}
\end{equation}
noting that the second term in the first line vanishes due to symmetry, and renaming contracted indices. We now see that in the sum of the last two equations, the Lagrange multiplier cancels, and we obtain the previously found connection field equations~\eqref{eq:affcongenfield}. Hence, the Lagrange multiplier approach yields the same field equations as the restricted variation, as expected.

\subsubsection{Metric teleparallel gravity}\label{sssec:metaffmet}
In the metric teleparallel case, we assume in addition to vanishing curvature also that the nonmetricity~\eqref{eq:affconnonmet} vanishes. In the restricted variation approach, it thus follows that also its variation
\begin{equation}
\delta Q_{\mu\nu\rho} = \nabla_{\mu}\delta g_{\nu\rho} - g_{\sigma\rho}\delta\Gamma^{\sigma}{}_{\nu\mu} - g_{\nu\sigma}\delta\Gamma^{\sigma}{}_{\rho\mu}
\end{equation}
must vanish. Together with the form~\eqref{eq:metaffflatvar} of the flatness preserving variation of the affine connection this reduces to
\begin{equation}
\delta Q_{\mu\nu\rho} = \nabla_{\mu}\delta g_{\nu\rho} - g_{\sigma\rho}\nabla_{\mu}\xi^{\sigma}{}_{\nu} - g_{\nu\sigma}\nabla_{\mu}\xi^{\sigma}{}_{\rho} = \nabla_{\mu}(\delta g_{\nu\rho} - 2\xi_{(\nu\rho)})\,,
\end{equation}
where we used the metric compatibility of the connection to commute lowering an index with the covariant derivative. Hence, we can set
\begin{equation}
\delta g_{\mu\nu} = 2\xi_{(\mu\nu)}\,,
\end{equation}
and consider the variation with respect to \(\xi_{\mu\nu}\) as the only necessary variation in order to derive the field equations. The total variation is then given by
\begin{equation}
\delta S = \int_M\left(\Theta^{\mu\nu}\xi_{(\mu\nu)} - W^{\mu\nu}\xi_{(\mu\nu)} - Y^{\mu\nu\rho}\nabla_{\rho}\xi_{\mu\nu}\right)\sqrt{-g}\dd^4x = \int_M\left(\Theta^{(\mu\nu)} - W^{(\mu\nu)} + \nabla_{\rho}Y^{\mu\nu\rho} - Y^{\mu\nu\rho}K^{\tau}{}_{\rho\tau}\right)\xi_{\mu\nu}\sqrt{-g}\dd^4x\,,
\end{equation}
where we have performed integration by parts, so that we find the field equations
\begin{equation}\label{eq:metaffmetfield}
W^{\mu\nu} - \nabla_{\rho}Y^{\mu\nu\rho} + Y^{\mu\nu\rho}T^{\tau}{}_{\tau\rho} = \Theta^{\mu\nu}\,,
\end{equation}
using the trace of the contortion tensor~\eqref{eq:contor}. Here we have omitted the symmetrization brackets on \(W^{\mu\nu}\) and \(\Theta^{\mu\nu}\), since these are symmetric by definition. It is instructive to decompose these field equations into their symmetric and antisymmetric components,
\begin{equation}\label{eq:metaffmetfieldas}
W^{\mu\nu} - \nabla_{\rho}Y^{(\mu\nu)\rho} + Y^{(\mu\nu)\rho}T^{\tau}{}_{\tau\rho} = \Theta^{\mu\nu}\,, \quad
\nabla_{\rho}Y^{[\mu\nu]\rho} - Y^{[\mu\nu]\rho}T^{\tau}{}_{\tau\rho} = 0\,,
\end{equation}
and to note that the antisymmetric part arises only from the variation with respect to the independent connection. Note that the symmetric and antisymmetric parts together have 16 independent components.

In order to obtain these field equations from the Lagrange multiplier method, one leaves the affine connection \(\Gamma^{\mu}{}_{\nu\rho}\) general a priori, and introduces a new term
\begin{equation}
S_{\text{l}} = \int_M(r_{\mu}{}^{\nu\rho\sigma}R^{\mu}{}_{\nu\rho\sigma} + q^{\mu(\nu\rho)}Q_{\mu\nu\rho})\sqrt{-g}\dd^4x
\end{equation}
into the total action. Variation with respect to the Lagrange multipliers \(r_{\mu}{}^{\nu\rho\sigma}\) and \(q^{\mu(\nu\rho)}\) then imposes the conditions of vanishing curvature~\eqref{eq:affconflat} and nonmetricity~\eqref{eq:affconnonmet}. By variation with respect to the now fully general connection coefficients one then obtains
\begin{equation}
\begin{split}
\delta_{\Gamma}S &= \int_M\left[r_{\mu}{}^{\nu\rho\sigma}\left(\nabla_{\rho}\delta\Gamma^{\mu}{}_{\nu\sigma} - \nabla_{\sigma}\delta\Gamma^{\mu}{}_{\nu\rho} + T^{\tau}{}_{\rho\sigma}\delta\Gamma^{\mu}{}_{\nu\tau}\right) - 2q^{\mu(\nu\rho)}g_{\sigma(\nu}\delta\Gamma^{\sigma}{}_{\rho)\mu} - Y_{\mu}{}^{\nu\rho}\delta\Gamma^{\mu}{}_{\nu\rho}\right]\sqrt{-g}\dd^4x\\
&= \int_M\left(2\nabla_{\sigma}r_{\mu}{}^{\nu[\rho\sigma]} - 2r_{\mu}{}^{\nu[\rho\sigma]}M^{\tau}{}_{\sigma\tau} - 2r_{\mu}{}^{\nu[\sigma\tau]}M^{\rho}{}_{\sigma\tau} - 2g_{\mu\sigma}q^{\rho(\nu\sigma)} - Y_{\mu}{}^{\nu\rho}\right)\delta\Gamma^{\mu}{}_{\nu\rho}\sqrt{-g}\dd^4x\,.
\end{split}
\end{equation}
Hence, we obtain the field equations
\begin{equation}\label{eq:metaffmetfieldl}
2\nabla_{\sigma}r^{\mu\nu[\rho\sigma]} - 2r^{\mu\nu[\rho\sigma]}K^{\tau}{}_{\sigma\tau} - 2r^{\mu\nu[\sigma\tau]}K^{\rho}{}_{\sigma\tau} - 2q^{\rho(\mu\nu)} - Y^{\mu\nu\rho} = 0\,,
\end{equation}
where we have raised one index for convenience, and replaced the distortion by the contortion, using the fact that the connection is metric as imposed by the Lagrange multiplier field equations. We see that the part of the equations which is symmetric in the indices \(\mu\) and \(\nu\) has the only role to fix the value of the Lagrange multipliers \(q^{\rho(\mu\nu)}\), but does not yield and restrictions on the dynamical fields. Hence, one may restrict the following considerations to the antisymmetric part
\begin{equation}\label{eq:metaffmetfieldla}
2\nabla_{\sigma}r^{[\mu\nu][\rho\sigma]} - 2r^{[\mu\nu][\rho\sigma]}K^{\tau}{}_{\sigma\tau} - 2r^{[\mu\nu][\sigma\tau]}K^{\rho}{}_{\sigma\tau} - Y^{[\mu\nu]\rho} = 0\,.
\end{equation}
In order to eliminate also the Lagrange multipliers \(r^{\mu\nu\rho\sigma}\) from the resulting equation, one may proceed in a similar fashion as for general teleparallel case discussed in the preceding section. We thus take the divergence with respect to the teleparallel connection \(\nabla_{\rho}\). Further using the relation
\begin{equation}
2\nabla_{\rho}\nabla_{\sigma}r^{[\mu\nu][\rho\sigma]} = 2\nabla_{[\rho}\nabla_{\sigma]}r^{[\mu\nu]\rho\sigma} = -T^{\tau}{}_{\rho\sigma}\nabla_{\tau}r^{[\mu\nu]\rho\sigma}\,,
\end{equation}
for the commutator of covariant derivatives, which follows from the fact that the teleparallel connection has vanishing curvature, but non-vanishing torsion, this leads to the equation
\begin{equation}
\nabla_{\rho}\left(Y^{[\mu\nu]\rho} + 2K^{\rho}{}_{\tau\sigma}r^{[\mu\nu][\tau\sigma]} + 2K^{\sigma}{}_{\tau\sigma}r^{[\mu\nu][\rho\tau]}\right) + T^{\tau}{}_{\rho\sigma}\nabla_{\tau}r^{[\mu\nu]\rho\sigma} = 0\,.
\end{equation}
We continue by expanding the contortion tensor in terms of its definition~\eqref{eq:contor}. After applying the product rule on the derivative terms, the equation becomes
\begin{equation}
\nabla_{\rho}Y^{[\mu\nu]\rho} - 2T^{\tau}{}_{\tau\rho}\nabla_{\sigma}r^{[\mu\nu][\rho\sigma]} - 3\nabla_{[\tau}T^{\tau}{}_{\rho\sigma]}r^{[\mu\nu][\rho\sigma]} = 0\,.
\end{equation}
On the last term, one can use the contracted Bianchi identity~\eqref{eq:bianchioneflatc}, which leads to the equation
\begin{equation}
\nabla_{\rho}Y^{[\mu\nu]\rho} - 2T^{\tau}{}_{\tau\rho}\nabla_{\sigma}r^{[\mu\nu][\rho\sigma]} - T^{\tau}{}_{\tau\omega}T^{\omega}{}_{\rho\sigma}r^{[\mu\nu][\rho\sigma]} = 0\,.
\end{equation}
We can compare this result with the antisymmetric equation~\eqref{eq:metaffmetfieldla}. One finds that it matches the covariant derivative of the Lagrange multiplier \(r^{[\mu\nu][\rho\sigma]}\), contracted with the trace \(T^{\tau}{}_{\tau\omega}\) of the torsion tensor. Using again the antisymmetric equation~\eqref{eq:metaffmetfieldla}, contracting it with the torsion in the same way, and finally expanding the contortion tensor~\eqref{eq:contor}, one obtains, after simplification, the equation
\begin{equation}
T^{\tau}{}_{\tau\omega}\left(2\nabla_{\sigma}r^{[\mu\nu][\omega\sigma]} + T^{\omega}{}_{\rho\sigma}r^{[\mu\nu][\rho\sigma]} - Y^{[\mu\nu]\omega}\right) = 0\,.
\end{equation}
It is now straightforward that by taking the sum of the last two equations, the Lagrange multiplier cancels. The remaining equation reproduces the antisymmetric part~\eqref{eq:metaffmetfieldas} of the field equations we derived using the restricted variation approach. In order to obtain also the symmetric part, we consider the variation of the total action with respect to the metric, which yields
\begin{equation}
\begin{split}
\delta_gS &= \int_M\left[\frac{1}{2}(\Theta^{\mu\nu} - W^{\mu\nu})\delta g_{\mu\nu} + q^{\rho(\mu\nu)}\nabla_{\rho}\delta g_{\mu\nu}\right]\sqrt{-g}\dd^4x\\
&= \frac{1}{2}\int_M\left(\Theta^{\mu\nu} - W^{\mu\nu} - 2\nabla_{\rho}q^{\rho(\mu\nu)} - 2q^{\sigma(\mu\nu)}M^{\rho}{}_{\sigma\rho}\right)\delta g_{\mu\nu}\sqrt{-g}\dd^4x
\end{split}
\end{equation}
after integration by parts. One therefore finds the field equations
\begin{equation}
W^{\mu\nu} + 2\nabla_{\rho}q^{\rho(\mu\nu)} + 2q^{\sigma(\mu\nu)}M^{\rho}{}_{\sigma\rho} = \Theta^{\mu\nu}\,.
\end{equation}
The Lagrange multiplier \(q^{\rho(\mu\nu)}\) can be substituted using the equation~\eqref{eq:metaffmetfieldl} obtained by variation with respect to the connection, leaving the Lagrange multiplier \(r^{(\mu\nu)[\rho\sigma]}\) instead. Repeating essentially the same calculation as we have done for the connection field equation, but for the symmetric part in the indices \(\mu\) and \(\nu\) instead of the antisymmetric part, one finds that the Lagrange multipliers cancel, and the metric field equation reduces to the symmetric part of the field equations~\eqref{eq:metaffmetfieldas}. Hence, we have obtained the same set of field equations as in the restricted variation approach.

\subsubsection{Symmetric teleparallel gravity}\label{sssec:metaffsym}
We finally come to the case of symmetric teleparallel gravity theories, where in addition to the curvature also the torsion~\eqref{eq:affcontors} of the independent affine connection is assumed to vanish. Using the approach of restricted variation, the corresponding torsion variation
\begin{equation}
\delta T^{\mu}{}_{\nu\rho} = \delta\Gamma^{\mu}{}_{\rho\nu} - \delta\Gamma^{\mu}{}_{\nu\rho}
\end{equation}
must therefore also vanish, and so the variation of the connection must be symmetric in its lower indices. Together with the form~\eqref{eq:metaffflatvar} of the variation of a flat affine connection, one thus has the condition
\begin{equation}
\nabla_{[\rho}\xi^{\mu}{}_{\nu]} = 0\,.
\end{equation}
These equations possess the general solution
\begin{equation}
\xi^{\mu}{}_{\nu} = \nabla_{\nu}\zeta^{\mu}
\end{equation}
in terms of a vector field \(\zeta^{\mu}\). Using the formula for the commutator of covariant derivatives, one immediately finds
\begin{equation}
2\nabla_{[\rho}\xi^{\mu}{}_{\nu]} = 2\nabla_{[\rho}\nabla_{\nu]}\zeta^{\mu} = R^{\mu}{}_{\sigma\rho\nu}\zeta^{\sigma} - T^{\sigma}{}_{\rho\nu}\nabla_{\sigma}\zeta^{\mu} = 0
\end{equation}
in the absence of curvature and torsion. For the variation of the action with respect to the curvature thus follows
\begin{equation}
\begin{split}
\delta_{\Gamma}S &= -\int_MY_{\mu}{}^{\nu\rho}\nabla_{\rho}\nabla_{\nu}\zeta^{\mu}\sqrt{-g}\dd^4x\\
&= \int_M(\nabla_{\rho}Y_{\mu}{}^{\nu\rho} - Y_{\mu}{}^{\nu\rho}L^{\tau}{}_{\rho\tau})\nabla_{\nu}\zeta^{\mu}\sqrt{-g}\dd^4x\\
&= -\int_M\left[\nabla_{\nu}(\nabla_{\rho}Y_{\mu}{}^{\nu\rho} - Y_{\mu}{}^{\nu\rho}L^{\tau}{}_{\rho\tau}) - (\nabla_{\rho}Y_{\mu}{}^{\nu\rho} - Y_{\mu}{}^{\nu\rho}L^{\tau}{}_{\rho\tau})L^{\omega}{}_{\nu\omega}\right]\zeta^{\mu}\sqrt{-g}\dd^4x\,,
\end{split}
\end{equation}
so that one obtains the field equation
\begin{equation}\label{eq:affconsymfield}
\nabla_{\nu}\nabla_{\rho}Y_{\mu}{}^{\nu\rho} + Q_{(\nu}\nabla_{\rho)}Y_{\mu}{}^{\nu\rho} + \frac{1}{2}\nabla_{(\nu}Q_{\rho)}Y_{\mu}{}^{\nu\rho} + \frac{1}{4}Q_{\nu}Q_{\rho}Y_{\mu}{}^{\nu\rho} = 0\,,
\end{equation}
using the shorthand notation \(Q_{\mu} = Q_{\mu\nu}{}^{\nu}\). A more convenient form of this equation can be found using the density~\eqref{eq:affconvardens}, which simply reads~\cite{BeltranJimenez:2018vdo}
\begin{equation}
\nabla_{\nu}\nabla_{\rho}\tilde{Y}_{\mu}{}^{\nu\rho} = 0\,.
\end{equation}
Finally, for the variation with respect to the metric, no restrictions are applied, and so one obtains the same metric field equation~\eqref{eq:metricgenfield} as in the general teleparallel case. Together with the four components of the connection equation one thus has 14 independent components.

We also derive this equation using the method of Lagrange multipliers. Here we add the term
\begin{equation}
S_{\text{l}} = \int_M(r_{\mu}{}^{\nu\rho\sigma}R^{\mu}{}_{\nu\rho\sigma} + t_{\mu}{}^{\nu\rho}T^{\mu}{}_{\nu\rho})\sqrt{-g}\dd^4x
\end{equation}
to the action, which implements the conditions of vanishing curvature and torsion. The variation of the total action with respect to the unrestricted affine connection then reads
\begin{equation}
\begin{split}
\delta_{\Gamma}S &= \int_M\left[r_{\mu}{}^{\nu\rho\sigma}\left(\nabla_{\rho}\delta\Gamma^{\mu}{}_{\nu\sigma} - \nabla_{\sigma}\delta\Gamma^{\mu}{}_{\nu\rho} + T^{\tau}{}_{\rho\sigma}\delta\Gamma^{\mu}{}_{\nu\tau}\right) - 2t_{\mu}{}^{\nu\rho}\delta\Gamma^{\mu}{}_{\nu\rho} - Y_{\mu}{}^{\nu\rho}\delta\Gamma^{\mu}{}_{\nu\rho}\right]\sqrt{-g}\dd^4x\\
&= \int_M\left(2\nabla_{\sigma}r_{\mu}{}^{\nu[\rho\sigma]} - 2r_{\mu}{}^{\nu[\rho\sigma]}M^{\tau}{}_{\sigma\tau} - 2r_{\mu}{}^{\nu[\sigma\tau]}M^{\rho}{}_{\sigma\tau} - 2t_{\mu}{}^{[\nu\rho]} - Y_{\mu}{}^{\nu\rho}\right)\delta\Gamma^{\mu}{}_{\nu\rho}\sqrt{-g}\dd^4x\,,
\end{split}
\end{equation}
and thus yields the field equations
\begin{equation}
2\nabla_{\sigma}r_{\mu}{}^{\nu[\rho\sigma]} - 2r_{\mu}{}^{\nu[\rho\sigma]}L^{\tau}{}_{\sigma\tau} - 2t_{\mu}{}^{[\nu\rho]} - Y_{\mu}{}^{\nu\rho} = 0\,,
\end{equation}
where we have replaced the distortion by the disformation, since the torsion vanishes, and omitted a term which cancels due to the symmetry of the disformation in its lower indices. We see that the part of the equations which is antisymmetric in the indices \(\nu\) and \(\rho\) simply determines the values of the Lagrange multipliers \(t_{\mu}{}^{[\nu\rho]}\), and can therefore be removed from the equations by retaining only the symmetric part, which reads
\begin{equation}\label{eq:lagmulsymfield}
Y_{\mu}{}^{(\nu\rho)} = \nabla_{\sigma}r_{\mu}{}^{\nu[\rho\sigma]} + \nabla_{\sigma}r_{\mu}{}^{\rho[\nu\sigma]} + \frac{1}{2}Q_{\sigma}(r_{\mu}{}^{\nu[\rho\sigma]} + r_{\mu}{}^{\rho[\nu\sigma]})\,,
\end{equation}
using the trace
\begin{equation}\label{eq:disfortrace}
L^{\mu}{}_{\mu\nu} = -\frac{1}{2}Q_{\nu\mu}{}^{\mu} = -\frac{1}{2}Q_{\nu}
\end{equation}
One then proceeds similarly to the calculation shown in section~\ref{sssec:metaffgen} in the case of general teleparallel gravity. In order to eliminate the remaining Lagrange multiplier \(r_{\mu}{}^{\nu[\rho\sigma]}\), one may attempt to calculate the divergence with respect to the covariant derivative \(\nabla_{\rho}\), which yields
\begin{equation}
\nabla_{\rho}Y_{\mu}{}^{(\nu\rho)} = \nabla_{\rho}\nabla_{\sigma}r_{\mu}{}^{\rho[\nu\sigma]} + \frac{1}{2}\nabla_{\rho}Q_{\sigma}r_{\mu}{}^{\rho[\nu\sigma]} + \frac{1}{2}Q_{\sigma}\nabla_{\rho}(r_{\mu}{}^{\nu[\rho\sigma]} + r_{\mu}{}^{\rho[\nu\sigma]})\,,
\end{equation}
where we have omitted terms which cancel due to symmetry. One realizes that the last terms are similar to contracting the original equation~\eqref{eq:lagmulsymfield} with the trace of the nonmetricity, which yields
\begin{equation}
Q_{\rho}Y_{\mu}{}^{(\nu\rho)} = Q_{\rho}\nabla_{\sigma}r_{\mu}{}^{\nu[\rho\sigma]} + Q_{\rho}\nabla_{\sigma}r_{\mu}{}^{\rho[\nu\sigma]} + \frac{1}{2}Q_{\rho}Q_{\sigma}r_{\mu}{}^{\rho[\nu\sigma]}\,,
\end{equation}
once again omitting a term which cancels due to symmetry. We see that in the combination
\begin{equation}
\nabla_{\rho}Y_{\mu}{}^{(\nu\rho)} + \frac{1}{2}Q_{\rho}Y_{\mu}{}^{(\nu\rho)} = \nabla_{\rho}\nabla_{\sigma}r_{\mu}{}^{\rho[\nu\sigma]} + \frac{1}{2}\nabla_{\rho}Q_{\sigma}r_{\mu}{}^{\rho[\nu\sigma]} + Q_{(\rho}\nabla_{\sigma)}r_{\mu}{}^{\rho[\nu\sigma]} + \frac{1}{4}Q_{\rho}Q_{\sigma}r_{\mu}{}^{\rho[\nu\sigma]}
\end{equation}
several terms cancel. To cancel the remaining terms, we once again take the divergence to obtain
\begin{equation}
\nabla_{\nu}\left(\nabla_{\rho}Y_{\mu}{}^{(\nu\rho)} + \frac{1}{2}Q_{\rho}Y_{\mu}{}^{(\nu\rho)}\right) = \frac{1}{2}Q_{\sigma}\nabla_{\nu}\nabla_{\rho}r_{\mu}{}^{\rho[\nu\sigma]} + \frac{1}{4}\nabla_{\nu}Q_{\rho}Q_{\sigma}r_{\mu}{}^{\rho[\nu\sigma]} + \frac{1}{4}Q_{\rho}Q_{\sigma}\nabla_{\nu}r_{\mu}{}^{\rho[\nu\sigma]}
\end{equation}
where we used
\begin{equation}
\nabla_{\rho}Q_{\sigma}\nabla_{\nu}r_{\mu}{}^{\rho[\nu\sigma]} + \nabla_{\nu}Q_{\rho}\nabla_{\sigma}r_{\mu}{}^{\rho[\nu\sigma]} = (\nabla_{\rho}Q_{\sigma} - \nabla_{\sigma}Q_{\rho})\nabla_{\nu}r_{\mu}{}^{\rho[\nu\sigma]} = 2\nabla_{[\rho}Q_{\sigma]}\nabla_{\nu}r_{\mu}{}^{\rho[\nu\sigma]} = 0\,,
\end{equation}
among other relations, to remove terms which vanish or cancel each other. We can compare this with the result obtained by contracting once again with the trace of the nonmetricity tensor, which reads
\begin{equation}
Q_{\nu}\left(\nabla_{\rho}Y_{\mu}{}^{(\nu\rho)} + \frac{1}{2}Q_{\rho}Y_{\mu}{}^{(\nu\rho)}\right) = Q_{\nu}\nabla_{\rho}\nabla_{\sigma}r_{\mu}{}^{\rho[\nu\sigma]} + \frac{1}{2}Q_{\nu}\nabla_{\rho}Q_{\sigma}r_{\mu}{}^{\rho[\nu\sigma]} + \frac{1}{2}Q_{\nu}Q_{\rho}\nabla_{\sigma}r_{\mu}{}^{\rho[\nu\sigma]}
\end{equation}
Taking into account the antisymmetrization in \(\nu\) and \(\sigma\) on the right hand side, we see that in the combined equation
\begin{equation}
\nabla_{\nu}\left(\nabla_{\rho}Y_{\mu}{}^{(\nu\rho)} + \frac{1}{2}Q_{\rho}Y_{\mu}{}^{(\nu\rho)}\right) + \frac{1}{2}Q_{\nu}\left(\nabla_{\rho}Y_{\mu}{}^{(\nu\rho)} + \frac{1}{2}Q_{\rho}Y_{\mu}{}^{(\nu\rho)}\right) = 0
\end{equation}
the Lagrange multiplier terms cancel. Finally, using the product rule and shifting the symmetrization brackets to the lower indices, we therefore re-obtain the connection field equation~\eqref{eq:affconsymfield}.

We conclude this section by pointing out a peculiar property of the field equations in the symmetric teleparallel class of gravity theories. From the diffeomorphism invariance condition~\eqref{eq:metaffdiffinvgrav} follows
\begin{equation}
\begin{split}
\lc{\nabla}_{\nu}W_{\mu}{}^{\nu} &= \lc{\nabla}_{\nu}\left(\nabla_{\rho}Y_{\mu}{}^{\nu\rho} - Y_{\mu}{}^{\nu\rho}L^{\sigma}{}_{\rho\sigma}\right) - \left(\nabla_{\sigma}Y_{\nu}{}^{\rho\sigma} - Y_{\nu}{}^{\rho\sigma}L^{\tau}{}_{\sigma\tau}\right)L^{\nu}{}_{\rho\mu}\\
&= \nabla_{\nu}\left(\nabla_{\rho}Y_{\mu}{}^{\nu\rho} - Y_{\mu}{}^{\nu\rho}L^{\sigma}{}_{\rho\sigma}\right) - \left(\nabla_{\rho}Y_{\mu}{}^{\sigma\rho} - Y_{\mu}{}^{\sigma\rho}L^{\tau}{}_{\rho\tau}\right)L^{\nu}{}_{\sigma\nu}\\
&= \nabla_{\nu}\nabla_{\rho}Y_{\mu}{}^{\nu\rho} + Q_{(\nu}\nabla_{\rho)}Y_{\mu}{}^{\nu\rho} + \frac{1}{2}\nabla_{(\nu}Q_{\rho)}Y_{\mu}{}^{\nu\rho} + \frac{1}{4}Q_{\nu}Q_{\rho}Y_{\mu}{}^{\nu\rho}\,,
\end{split}
\end{equation}
where we used that \(M^{\mu}{}_{\nu\rho} = L^{\mu}{}_{\nu\rho}\) in the absence of torsion, the symmetry \(L^{\mu}{}_{[\nu\rho]} = 0\) of the disformation, its trace~\eqref{eq:disfortrace} and the commutativity~\eqref{eq:nonmetasymder} of the symmetric teleparallel covariant derivative, \(\nabla_{[\mu}Q_{\nu]} = 0\). We see that the right hand side is simply the connection equation~\eqref{eq:affconsymfield}, whose components are therefore not independent of the metric field equations. Hence, we find that for any solution of the metric field equation~\eqref{eq:metricgenfield}, which satisfies \(\lc{\nabla}_{\nu}W_{\mu}{}^{\nu} = 0\) as a consequence of the matter energy-momentum conservation~\eqref{eq:metaffdiffinvmat}, also the connection field equation~\eqref{eq:affconsymfield} is solved~\cite{BeltranJimenez:2018vdo}.

\subsection{Tetrad formulation}\label{ssec:tetradgrav}
Having derived the field equations for all three classes of teleparallel gravity theories in the metric-affine formulation, we continue with the tetrad formulation. As in the metric-affine case, we start with a number of general remarks in section~\ref{sssec:tetradvar}. These are then applied to the different classes of teleparallel gravity theories. In particular, we consider general teleparallel gravity in section~\ref{sssec:tetradgen}, metric teleparallel gravity in section~\ref{sssec:tetradmet} and symmetric teleparallel gravity in section~\ref{sssec:tetradsym}.

\subsubsection{General action and variation}\label{sssec:tetradvar}
We begin our discussion of teleparallel gravity theories in the tetrad formulation by providing the necessary definitions and notation. Similarly to the metric-affine formulation, we split the action in the form
\begin{equation}\label{eq:tetradactsplit}
S = S_{\text{g}}[\theta, \omega] + S_{\text{m}}[\theta, \psi] + S_{\text{l}}[\theta, \omega, r, t, q]\,,
\end{equation}
where the three parts \(S_{\text{g}}, S_{\text{m}}, S_{\text{l}}\) have the same meaning as in the split~\eqref{eq:metaffactsplit}. Also here we assume that the matter action does not depend on the connection variable, which in this case is the spin connection \(\omega^a{}_b\), and is the only part which involves the matter fields \(\psi^I\), which we now assume to be given by arbitrary differential forms. It follows that we may write the general variation of the matter action as
\begin{equation}\label{eq:tetradmatactvar}
\delta S_{\text{m}} = \int_M(\Sigma_a \wedge \delta\theta^a + \Psi_I \wedge \delta\psi^I)\,,
\end{equation}
where we have introduced the energy-momentum three-forms \(\Sigma_a\), as well as differential forms \(\Psi_I\), whose rank depends on the rank of the matter fields \(\psi^I\), and which constitute the matter field equations \(\Psi_I = 0\). Similarly, we write the variation of the gravitational part of the action in the form
\begin{equation}\label{eq:tetradgravactvar}
\delta S_{\text{g}} = \int_M(\Delta_a \wedge \delta\theta^a + \Xi_a{}^b \wedge \delta\omega^a{}_b)\,,
\end{equation}
where \(\Delta_a\) and \(\Xi_a{}^b\) are three-forms.

In order to achieve an equivalence between the tetrad and metric-affine formulations of teleparallel gravity theories, we must further demand that each component of the action~\eqref{eq:tetradactsplit} is invariant under local Lorentz transformations~\eqref{eq:finloclortrans} of the tetrad and the spin connection. For the matter action, this implies that the variation of the action with respect to an infinitesimal local Lorentz transformation
\begin{equation}\label{eq:infloclortrans}
\delta_{\lambda}\theta^a = \lambda^a{}_b\theta^b\,, \quad
\delta_{\lambda}\omega^a{}_b = \lambda^a{}_c\omega^c{}_b - \omega^a{}_c\lambda^c{}_b - \dd\lambda^a{}_b = -\DD\lambda^a{}_b\,,
\end{equation}
which follows from the finite case~\eqref{eq:finloclortrans} with \(\Lambda^a{}_b = \delta^a_b + \lambda^a{}_b\) and \(\lambda_{(ab)} = 0\), vanishes. Hence,
\begin{equation}\label{eq:lorinvmatact}
0 = \delta_{\lambda}S_{\text{m}} = \int_M\Sigma_a \wedge (\lambda^a{}_b\theta^b) = \int_M\Sigma^{[a} \wedge \theta^{b]}\lambda_{ab}\,,
\end{equation}
since only the antisymmetric part \(\lambda_{[ab]}\) contributes. This vanishes for arbitrary local Lorentz transformations if and only if the energy-momentum three-forms are symmetric,
\begin{equation}\label{eq:lorinvmat}
\Sigma^{[a} \wedge \theta^{b]} = 0\,.
\end{equation}
For the gravitational part of the action, one analogously demands the Lorentz invariance
\begin{equation}\label{eq:lorinvgravact}
0 = \delta_{\lambda}S_{\text{g}} = \int_M\left[\Delta_a \wedge (\lambda^a{}_b\theta^b) - \Xi_a{}^b \wedge \DD\lambda^a{}_b\right] = \int_M\left(\Delta^{[a} \wedge \theta^{b]} - \eta^{c[a}\DD\Xi_c{}^{b]}\right)\lambda_{ab}\,,
\end{equation}
where we have performed integration by parts in the second term. It thus follows that the condition for local Lorentz invariance reads
\begin{equation}\label{eq:lorinvgrav}
\Delta^{[a} \wedge \theta^{b]} - \eta^{c[a}\DD\Xi_c{}^{b]} = 0\,.
\end{equation}
Note that, in general, one cannot commute raising one Lorentz index and taking the exterior covariant derivative, since in the presence on nonmetricity one has \(\DD\eta^{ab} \neq 0\). Finally, we must also impose local Lorentz invariance of the Lagrange multiplier contribution \(S_{\text{l}}\). This will be done later, by explicitly specifying a Lorentz invariant action.

As in the metric-affine approach, we also here demand that each part of the action is separately invariant under diffeomorphisms. Under an infinitesimal diffeomorphism generated by a vector field \(X^{\mu}\), the tetrad and spin connection transform as one-forms satisfying Cartan's formula
\begin{equation}
\delta\theta^a = \mathcal{L}_X\theta^a = \dd(X \intprod \theta^a) + X \intprod \dd\theta^a\,, \quad
\delta\omega^a{}_b = \mathcal{L}_X\omega^a{}_b = \dd(X \intprod \omega^a{}_b) + X \intprod \dd\omega^a{}_b\,.
\end{equation}
For the matter action, the diffeomorphism invariance implies that on-shell, when the matter field equations \(\Psi_I = 0\) are satisfied, the energy-momentum three-form satisfies~\cite{Hohmann:2018vle}
\begin{equation}
\begin{split}
0 = \delta_XS_{\text{m}} &= \int_M\{\Sigma_a \wedge [\dd(X \intprod \theta^a) + X \intprod \dd\theta^a]\}\\
&= \int_M[\dd\Sigma_a + \Sigma_b \wedge (e_a \intprod \dd\theta^b)](X \intprod \theta^a)\\
&= \int_M[\dd\Sigma_a + \Sigma_b \wedge \lc{\omega}^b{}_a](X \intprod \theta^a)\\
&= \int_M\lc{\DD}\Sigma_a(X \intprod \theta^a)\,,
\end{split}
\end{equation}
where we used the Lorentz invariance condition~\eqref{eq:lorinvmat} to cancel the symmetric contribution from the Levi-Civita connection~\eqref{eq:levicivitaspi}. Since this it imposed for arbitrary vector fields \(X^{\mu}\), one finds
\begin{equation}\label{eq:tetraddiffinvmat}
\lc{\DD}\Sigma_a = 0\,,
\end{equation}
and so the energy-momentum three-form is conserved. For the gravitational part of the action, one similarly finds
\begin{equation}
\begin{split}
0 = \delta_XS_{\text{g}} &= \int_M(\Delta_a \wedge \mathcal{L}_X\theta^a + \Xi_a{}^b \wedge \mathcal{L}_X\omega^a{}_b)\\
&= \int_M\{\Delta_a \wedge [\dd(X \intprod \theta^a) + X \intprod \dd\theta^a] + \Xi_a{}^b \wedge [\dd(X \intprod \omega^a{}_b) + X \intprod \dd\omega^a{}_b]\}\\
&= \int_M[\dd\Delta_a + \Delta_b \wedge (e_a \intprod \dd\theta^b) + \dd\Xi_b{}^c \wedge (e_a \intprod \omega^b{}_c) + \Xi_b{}^c \wedge (e_a \intprod \dd\omega^b{}_c)](X \intprod \theta^a)\,,
\end{split}
\end{equation}
and so one obtains the condition
\begin{equation}\label{eq:diffinvspivar}
\dd\Delta_a + \Delta_b \wedge (e_a \intprod \dd\theta^b) + \dd\Xi_b{}^c \wedge (e_a \intprod \omega^b{}_c) + \Xi_b{}^c \wedge (e_a \intprod \dd\omega^b{}_c) = 0\,.
\end{equation}
This can be simplified in a few steps. Using the flatness~\eqref{eq:spiconflat} of the teleparallel connection, one can replace \(\dd\omega^b{}_c\) in the last term, and so the contribution from \(\Xi_a{}^b\) becomes
\begin{equation}
\dd\Xi_b{}^c \wedge (e_a \intprod \omega^b{}_c) + \Xi_b{}^c \wedge (e_a \intprod \dd\omega^b{}_c) = \dd\Xi_b{}^c \wedge (e_a \intprod \omega^b{}_c) - \Xi_b{}^c \wedge [e_a \intprod (\omega^b{}_d \wedge \omega^d{}_c)] = \DD\Xi_b{}^c \wedge (e_a \intprod \omega^b{}_c)\,.
\end{equation}
In the next step, it is helpful to apply the decomposition~\eqref{eq:spicondec}. Using the fact that the Levi-Civita spin connection is antisymmetric, one then has
\begin{equation}
\DD\Xi_b{}^c \wedge (e_a \intprod \omega^b{}_c) = \eta^{d[b}\DD\Xi_d{}^{c]} \wedge (e_a \intprod \lc{\omega}_{bc}) + \DD\Xi_b{}^c \wedge (e_a \intprod M^b{}_c)\,.
\end{equation}
We then turn our attention to the first term. Using the Lorentz invariance condition~\eqref{eq:lorinvgrav} and the definition~\eqref{eq:levicivitaspi} this can be written as
\begin{equation}
\eta^{d[b}\DD\Xi_d{}^{c]} \wedge (e_a \intprod \lc{\omega}_{bc}) = -\frac{1}{2}\Delta^{[b} \wedge \theta^{c]} \wedge \left(e_c \intprod e_a \intprod \dd\theta_b + e_a \intprod e_b \intprod \dd\theta_c - e_b \intprod e_c \intprod \dd\theta_a\right)\,,
\end{equation}
and one can omit the antisymmetrization on both sides, since the Levi-Civita connection is already antisymmetric by definition. In combination with the second term from the original equation~\eqref{eq:diffinvspivar}, one thus has
\begin{equation}
\Delta_b \wedge (e_a \intprod \dd\theta^b) + \eta^{d[b}\DD\Xi_d{}^{c]} \wedge (e_a \intprod \lc{\omega}_{bc}) = -\Delta^b \wedge \lc{\omega}_{ab} = -\lc{\omega}^b{}_a \wedge \Delta_b\,,
\end{equation}
which is further combined with the first term to yield a covariant exterior derivative. Thus, the condition~\eqref{eq:diffinvspivar} is finally rewritten as
\begin{equation}\label{eq:tetraddiffinvgrav}
\lc{\DD}\Delta_a + \DD\Xi_b{}^c \wedge (e_a \intprod M^b{}_c) = 0\,.
\end{equation}
Note that this derivation holds independently of any gravitational field equations, and so the resulting equation it must be a geometric identity, which is satisfied by any teleparallel geometry, irrespective of whether it satisfies the field equations or not.

\subsubsection{General teleparallel gravity}\label{sssec:tetradgen}
Again in analogy to the metric-affine formulation, we now derive the field equations for the general teleparallel class of gravity theories, in which the connection must satisfy only the flatness condition~\eqref{eq:spiconflat}, but may have both non-vanishing torsion and nonmetricity. Also here we follow two approaches, by either restricting the variation of the spin connection such that it remains flat, or by imposing the flatness via a suitable Lagrange multiplier. Starting with the former approach, note that the variation of the curvature is given by
\begin{equation}\label{eq:spiconvarflat}
\delta R^a{}_b = \dd\delta\omega^a{}_b + \delta\omega^a{}_c \wedge \omega^c{}_b + \omega^a{}_c \wedge \delta\omega^c{}_b = \DD\delta\omega^a{}_b\,.
\end{equation}
In order for this to vanish, one sets
\begin{equation}
\delta\omega^a{}_b = \DD\xi^a{}_b
\end{equation}
with arbitrary zero-forms \(\xi^a{}_b\), or equivalently
\begin{equation}
\delta\Omega^a{}_b = \Omega^a{}_c\xi^c{}_b
\end{equation}
using the integral form~\eqref{eq:spiconinteg} of the spin connection. From this follows
\begin{equation}
\delta R^a{}_b = \DD^2\xi^a{}_b = R^a{}_c \wedge \xi^c{}_b - R^c{}_b \wedge \xi^a{}_c = 0\,,
\end{equation}
due to the vanishing curvature. The connection field equation then follows from the variation
\begin{equation}\label{eq:spiconflatactvar}
0 = \delta_{\omega}S = \int_M\Xi_a{}^b \wedge \DD\xi^a{}_b = \int_M\DD\Xi_a{}^b \wedge \xi^a{}_b\,,
\end{equation}
and so reads
\begin{equation}\label{eq:spicongenfield}
\DD\Xi_a{}^b = 0\,.
\end{equation}
For the tetrad, the variation takes the form
\begin{equation}\label{eq:tetradactvar}
0 = \delta_{\theta}S = \int_M(\Delta_a + \Sigma_a) \wedge \delta\theta^a\,,
\end{equation}
and so the field equations simply read
\begin{equation}\label{eq:tetradgenfield}
\Delta_a + \Sigma_a = 0\,.
\end{equation}
It is worth taking a closer look at the antisymmetric part of the field equations. Due to the symmetry~\eqref{eq:lorinvmat} of the energy-momentum three-forms, these read
\begin{equation}\label{eq:tetradgenfielda}
\Delta^{[a} \wedge \theta^{b]} = 0\,,
\end{equation}
and because of the Lorentz invariance~\eqref{eq:lorinvgrav}, are thus identical to the antisymmetric part of the spin connection field equations. The number of independent field equations is thus 10 from the symmetric part of the tetrad field equations, 10 from the symmetric part of the connection field equations, and 6 from the common antisymmetric part of both equations.

This approach can be contrasted with the Lagrange multiplier method. In this case the spin connection \(\omega^a{}_b\) and its variation \(\delta\omega^a{}_b\) are not a priori restricted to obey the flatness condition, but this is imposed by adding a Lagrange multiplier term
\begin{equation}\label{eq:spiconflatlag}
S_{\text{l}} = \int_Mr_a{}^b \wedge R^a{}_b
\end{equation}
to the action, with the Lagrange multiplier two-forms \(r_a{}^b\). Variation with respect to the Lagrange multiplier then yields the flatness condition~\eqref{eq:spiconflat}, while the variation with respect to the tetrad is unaffected. However, for the spin connection one now considers an arbitrary variation, and so obtains
\begin{equation}
0 = \delta_{\omega}S = \int_M\left(\Xi_a{}^b \wedge \delta\omega^a{}_b + r_a{}^b \wedge \DD\delta\omega^a{}_b\right) = \int_M\left(\Xi_a{}^b - \DD r_a{}^b\right) \wedge \delta\omega^a{}_b\,,
\end{equation}
so that one arrives at the field equations
\begin{equation}
\Xi_a{}^b - \DD r_a{}^b = 0\,,
\end{equation}
which now involve also the Lagrange multipliers. In order to eliminate these from the field equations, one makes use of the flatness of the connection, which implies \(\DD^2 = 0\), by taking the covariant derivative to obtain
\begin{equation}
0 = \DD\Xi_a{}^b - \DD^2r_a{}^b = \DD\Xi_a{}^b\,,
\end{equation}
and so one finds the same equations for the spin connection as with the restricted variation approach, hence proving the equivalence of both approaches.

\subsubsection{Metric teleparallel gravity}\label{sssec:tetradmet}
We continue with the metric teleparallel class of gravity theories, where in addition to the flatness we also impose vanishing nonmetricity~\eqref{eq:spiconnonmet}. Following the restricted variation approach, we therefore introduce the additional restriction that the variation
\begin{equation}
\delta Q_{ab} = -2\delta\omega_{(ab)} = -2\DD\xi_{(ab)}
\end{equation}
of the nonmetricity vanishes, and so the variation parameters \(\xi_{ab}\) must be antisymmetric in their indices, \(\xi_{(ab)} = 0\). It follows that the corresponding variation~\eqref{eq:spiconflatactvar} of the action now reads
\begin{equation}
0 = \delta_{\omega}S = \int_M\Xi^{[ab]} \wedge \DD\xi_{ab} = \int_M\DD\Xi^{[ab]} \wedge \xi_{ab}\,,
\end{equation}
so that the spin connection field equations become
\begin{equation}\label{eq:spiconmetfield}
\DD\Xi^{[ab]} = 0\,.
\end{equation}
For the variation with respect to the tetrad, nothing changes compared to the general teleparallel case we discussed before; the variation~\eqref{eq:tetradactvar} leads to the field equations~\eqref{eq:tetradgenfield}. In this case we find that the antisymmetric part~\eqref{eq:tetradgenfielda} of the latter is identical to the spin connection field equations, and so the latter do not constitute an independent field equation. Hence, there are only 10 + 6 independent components of the field equations. The latter reflects the fact that the spin connection, which is constrained by the conditions of vanishing curvature and non to be of the form~\eqref{eq:spiconinteg} with \(\Omega^a{}_b\) given by a local Lorentz transformation, can always be chosen to vanish by performing the inverse Lorentz transformation on both the tetrad and the spin connection. Hence, it is not a physical degree of freedom, and thus cannot mediate the gravitational interaction; the latter is attributed to the tetrad only. Instead, the role of the spin connection is merely to implement the local Lorentz invariance of the metric teleparallel gravity, which otherwise appears broken in a pure-tetrad formulation, due to the explicit Lorentz gauge choice with vanishing spin connection in this approach, as discussed in~\cite{Golovnev:2017dox}.

Following the Lagrange multiplier approach, we introduce another set of Lagrange multiplier three-forms \(q^{ab} = q^{(ab)}\), thus enhancing the corresponding term~\eqref{eq:spiconflatlag} to read
\begin{equation}
S_{\text{l}} = \int_M(r_a{}^b \wedge R^a{}_b + q^{(ab)} \wedge Q_{ab})\,.
\end{equation}
Variation with respect to the Lagrange multipliers then imposes the conditions of vanishing curvature and nonmetricity, while variation with respect to the spin connection now reads
\begin{equation}
0 = \delta_{\omega}S = \int_M\left(\Xi_a{}^b \wedge \delta\omega^a{}_b + r_a{}^b \wedge \DD\delta\omega^a{}_b - 2q^{ab} \wedge \delta\omega_{(ab)}\right) = \int_M\left(\Xi^{ab} - \DD r^{ab} - 2q^{(ab)}\right) \wedge \delta\omega_{ab}\,,
\end{equation}
which yields the field equation
\begin{equation}
\Xi^{ab} - \DD r^{ab} - 2q^{(ab)} = 0\,.
\end{equation}
The symmetric part of this equation simply determines the value of the Lagrange multiplier \(q^{(ab)}\), and can therefore be omitted as a field equation for the physical fields. One is thus left with the antisymmetric part
\begin{equation}
\Xi^{[ab]} - \DD r^{[ab]} = 0\,.
\end{equation}
Finally, one proceeds as in the general teleparallel case and uses the flatness \(\DD^2 = 0\) to eliminate also the Lagrange multiplier \(r^{[ab]}\), so that the remaining field equation becomes the same as in the restricted variation approach.

\subsubsection{Symmetric teleparallel gravity}\label{sssec:tetradsym}
Finally, we discuss the class of symmetric teleparallel gravity theories, in which one assumes vanishing curvature and torsion, but allows for non-vanishing nonmetricity. In the restricted variation approach, one must therefore implement the condition that the variation
\begin{equation}
\delta T^a = \DD\delta\theta^a + \delta\omega^a{}_b \wedge \theta^b = \DD\delta\theta^a + \DD\xi^a{}_b \wedge \theta^b
\end{equation}
of the torsion vanishes. This can be done by making the ansatz
\begin{equation}
\xi^a{}_b = e_b \intprod (\DD\zeta^a - \delta\theta^a)\,,
\end{equation}
in terms of zero-forms \(\zeta^a\). By direct calculation, using the product rule for the exterior covariant derivative as well as the vanishing curvature~\eqref{eq:spiconflat} and torsion~\eqref{eq:spicontors}, one easily shows that indeed
\begin{equation}
\begin{split}
\delta T^a &= \DD\delta\theta^a + \DD\left[e_b \intprod (\DD\zeta^a - \delta\theta^a)\right] \wedge \theta^b\\
&= \DD\delta\theta^a + \DD\left\{\left[e_b \intprod (\DD\zeta^a - \delta\theta^a)\right] \wedge \theta^b\right\} - \left[e_b \intprod (\DD\zeta^a - \delta\theta^a)\right] \wedge \DD\theta^b\\
&= \DD\delta\theta^a + \DD(\DD\zeta^a - \delta\theta^a) - \left[e_b \intprod (\DD\zeta^a - \delta\theta^a)\right] \wedge \DD\theta^b\\
&= \DD\delta\theta^a + R^a{}_b \wedge \zeta^b - \DD\delta\theta^a - \left[e_b \intprod (\DD\zeta^a - \delta\theta^a)\right] \wedge T^b\\
&= 0\,.
\end{split}
\end{equation}
Since \(\zeta^a\) and \(\delta\theta^a\) are arbitrary, independent variations, we can consider them separately, starting with the case \(\delta\theta^a = 0\). Then the variation of the action with respect to \(\zeta^a\) alone is given by
\begin{equation}
\delta_{\zeta}S = \int_M\Xi_a{}^b \wedge \DD(e_b \intprod \DD\zeta^a) = -\int_M\DD(e_b \intprod \DD\Xi_a{}^b) \wedge \zeta^a\,,
\end{equation}
and so we find the corresponding field equation
\begin{equation}\label{eq:spiconsymfield}
\DD(e_b \intprod \DD\Xi_a{}^b) = 0\,.
\end{equation}
We then continue with the tetrad variation, where we must take into account that this incurs also a term resulting from the dependent variation of the spin connection. Hence, the variation of the total action reads
\begin{equation}
0 = \delta_{\theta}S = \int_M\left[(\Delta_a + \Sigma_a) \wedge \delta\theta^a - \Xi_a{}^b \wedge \DD(e_b \intprod \delta\theta^a)\right] = \int_M(\Delta_a + \Sigma_a + e_b \intprod \DD\Xi_a{}^b) \wedge \delta\theta^a\,
\end{equation}
which results in the field equations
\begin{equation}\label{eq:tetradsymfield}
\Delta_a + \Sigma_a + e_b \intprod \DD\Xi_a{}^b = 0\,.
\end{equation}
Note that the spin connection field equation~\eqref{eq:spiconsymfield} is a four-form with a free Lorentz index, and so has four independent components, while the tetrad equations~\eqref{eq:tetradsymfield} are three-forms with the same number of Lorentz indices, and so yield 16 equations. Taking into account the Lorentz invariance, as a consequence of which six equations are not independent, one arrives at 14 independent equations.

For comparison, we also derive the field equations using the Lagrange multiplier method. Here we introduce a term
\begin{equation}
S_{\text{l}} = \int_M(r_a{}^b \wedge R^a{}_b + t_a \wedge T^a)
\end{equation}
into the action, where the Lagrange multiplier two-forms \(r_a{}^b\) and \(t_a\) impose the conditions of vanishing curvature and torsion. The variation of the total action with respect to an arbitrary, unrestricted spin connection is then given by
\begin{equation}
0 = \delta_{\omega}S = \int_M\left(\Xi_a{}^b \wedge \delta\omega^a{}_b + r_a{}^b \wedge \DD\delta\omega^a{}_b + t_a \wedge \delta\omega^a{}_b \wedge \theta^b\right) = \int_M\left(\Xi_a{}^b - \DD r_a{}^b - t_a \wedge \theta^b\right) \wedge \delta\omega^a{}_b\,.
\end{equation}
Hence, we find the field equations
\begin{equation}
\Xi_a{}^b - \DD r_a{}^b - t_a \wedge \theta^b = 0\,,
\end{equation}
from which the Lagrange multipliers must be eliminated. We see that \(r_a{}^b\) contributes to the field equations in form of a total derivative, which is eliminated by applying once again the exterior covariant derivative \(\DD\), so that one obtains
\begin{equation}
\DD\Xi_a{}^b - \DD t_a \wedge \theta^b = 0\,,
\end{equation}
and we omitted a vanishing torsion term \(\DD\theta^b\). In the next step, we take the interior product with \(e_b\), using the relations
\begin{equation}
(e_b \intprod \DD t_a) \wedge \theta^b = 3\DD t_a\,, \quad
e_b \intprod \theta^b = \delta^b_b = 4
\end{equation}
for the three-form \(\DD t_a\). Hence, we obtain the equation
\begin{equation}\label{eq:lagsymfield}
e_b \intprod \DD\Xi_a{}^b + \DD t_a = 0\,.
\end{equation}
Finally, applying once more the exterior covariant derivative \(\DD\), the Lagrange multiplier term \(\DD^2t_a = 0\) drops out and one is left with the previously found field equation~\eqref{eq:spiconsymfield}. Proceeding analogously with the tetrad variation, we find that the variation of the full action is given by
\begin{equation}
0 = \delta_{\theta}S = \int_M(\Delta_a \wedge \delta\theta^a + \Sigma_a \wedge \delta\theta^a + t_a \wedge \DD\delta\theta^a) = \int_M(\Delta_a + \Sigma_a - \DD t_a) \wedge \delta\theta^a\,,
\end{equation}
and hence leads to the field equations
\begin{equation}
\Delta_a + \Sigma_a - \DD t_a\,.
\end{equation}
Here we are left with the Lagrange multiplier term \(\DD t_a\), which must be eliminated. However, note that it is not sufficient to take the exterior covariant derivative of this equation, since it is not independent of the previously derived equation~\eqref{eq:lagsymfield}, which also involves \(\DD t_a\). Combining these equations, one finally arrives at the field equation~\eqref{eq:tetradsymfield}, which we have also obtained from the restricted variation approach, thus showing the equivalence of both approaches.

We finally return to the remark made at the end of section~\ref{sssec:metaffsym}, where we have shown that the connection field equations can be derived from the metric field equations and the diffeomorphism invariance. This can also be shown in the tetrad formulation. Taking the Levi-Civita exterior covariant derivative of the tetrad field equation~\eqref{eq:tetradsymfield}, and applying the diffeomorphism invariance~\eqref{eq:tetraddiffinvgrav}, one finds
\begin{equation}
0 = \lc{\DD}\left(\Delta_a + \Sigma_a + e_b \intprod \DD\Xi_a{}^b\right) = \lc{\DD}(e_b \intprod \DD\Xi_a{}^b) + (e_a \intprod \DD\Xi_b{}^c) \wedge M^b{}_c = \DD(e_b \intprod \DD\Xi_a{}^b)\,,
\end{equation}
which is the spin connection field equation~\eqref{eq:spiconsymfield}. Hence, any solution of the tetrad field equation also satisfies the spin connection field equation, as a consequence of the diffeomorphism invariance of the action.

\subsection{Relation between metric-affine and tetrad formulations}\label{ssec:gravrel}
In the previous sections we have derived the field equations for the different classes of teleparallel gravity theories both in the metric-affine and tetrad formulation, using the methods of Lagrange multipliers and constrained variation. We will now show that the two geometric formulations we used are indeed equivalent. As with the previous sections, we proceed in several steps. We start by discussing the relation between the variations of the fundamental fields in the different geometric frameworks and its consequences for the matter action in section~\ref{sssec:relvar}, before we apply these findings to the different classes of teleparallel gravity theories. These are general teleparallel gravity in section~\ref{sssec:relgen}, metric teleparallel gravity in section~\ref{sssec:relmet} and symmetric teleparallel gravity in section~\ref{sssec:relsym}.

\subsubsection{General action and variation}\label{sssec:relvar}
In order to relate the metric-affine and tetrad formulations, one needs to relate the variations of the different geometries to each other. This relation follows from the relations~\eqref{eq:metric} and~\eqref{eq:affcon} defining the metric and affine connection in terms of the tetrad and spin connection. Variation with respect to the latter yields the metric variation
\begin{equation}\label{eq:metricvar}
\delta g_{\mu\nu} = 2\eta_{ab}\theta^a{}_{(\mu}\delta\theta^b{}_{\nu)}
\end{equation}
and the affine connection
\begin{equation}\label{eq:affconvar}
\delta\Gamma^{\mu}{}_{\nu\rho} = e_a{}^{\mu}\left(\partial_{\rho}\delta\theta^a{}_{\nu} + \omega^a{}_{b\rho}\delta\theta^b{}_{\nu} - \Gamma^{\sigma}{}_{\nu\rho}\delta\theta^a{}_{\sigma} + \delta\omega^a{}_{b\rho}\theta^b{}_{\nu}\right)\,.
\end{equation}
In particular, it follows from the invariance of the metric and affine connection under local Lorentz transformations~\eqref{eq:finloclortrans} of the tetrad and spin connection that their variations vanish if they are induced by an infinitesimal Lorentz transformation~\eqref{eq:infloclortrans}. It follows that starting from any teleparallel gravity action in the metric-affine formulation and expressing it in terms of the tetrad formulation by using the relations~\eqref{eq:metric} and~\eqref{eq:affcon} leads to an action which satisfies the conditions~\eqref{eq:lorinvmatact} and~\eqref{eq:lorinvgravact} of local Lorentz invariance, which is the reason for imposing these conditions in the preceding section. For the matter action, it follows that the variation~\eqref{eq:metricmatactvar} of the matter action can be expressed as
\begin{equation}
\delta S_{\text{m}} = \int_M\left(\eta_{ab}\Theta^{(\mu\nu)}\theta^a{}_{\mu}\delta\theta^b{}_{\nu} + \Psi_I\delta\psi^I\right)\,\theta\,\dd^4x\,,
\end{equation}
while the variation~\eqref{eq:metricgravactvar} of the gravitational action becomes
\begin{equation}
\begin{split}
\delta S_{\text{g}} &= -\int_M\left[\eta_{ab}W^{(\mu\nu)}\theta^a{}_{\mu}\delta\theta^b{}_{\nu} + Y_{\mu}{}^{\nu\rho}e_a{}^{\mu}\left(\partial_{\rho}\delta\theta^a{}_{\nu} + \omega^a{}_{b\rho}\delta\theta^b{}_{\nu} - \Gamma^{\sigma}{}_{\nu\rho}\delta\theta^a{}_{\sigma} + \delta\omega^a{}_{b\rho}\theta^b{}_{\nu}\right)\right]\,\theta\,\dd^4x\\
&= -\int_M\left[\left(W_{\mu}{}^{\nu} - \nabla_{\rho}Y_{\mu}{}^{\nu\rho} + Y_{\mu}{}^{\nu\sigma}M^{\rho}{}_{\sigma\rho}\right)e_a{}^{\mu}\delta\theta^a{}_{\nu} + Y_{\mu}{}^{\nu\rho}e_a{}^{\mu}\theta^b{}_{\nu}\delta\omega^a{}_{b\rho}\right]\,\theta\,\dd^4x\,,
\end{split}
\end{equation}
where we have performed integration by parts, replaced the metric determinant \(\sqrt{-g}\) by the tetrad determinant \(\theta\) and omitted the symmetrization brackets on the already symmetric term \(W^{\mu\nu}\). In order to relate these expressions to the differential form language we used in the preceding section, it is helpful to write the scalar product of two one-forms induced by the metric in terms of the hodge star, which reads
\begin{equation}
\beta \wedge \star\alpha = \alpha \wedge \star\beta = \langle\alpha, \beta\rangle\,\mathrm{vol}_{\theta} = g^{\mu\nu}\alpha_{\mu}\beta_{\nu}\sqrt{-g}\dd^4x\,.
\end{equation}
Together with the hodge dual
\begin{equation}
\star\theta^a = \eta^{ab}e_b \intprod \mathrm{vol}_{\theta} = \frac{1}{6}\eta^{ab}\epsilon_{bcde}\theta^c \wedge \theta^d \wedge \theta^e = \frac{1}{6}\theta^a{}_{\tau}g^{\tau\mu}\epsilon_{\mu\nu\rho\sigma}\dd x^{\nu} \wedge \dd x^{\rho} \wedge \dd x^{\sigma}
\end{equation}
of the tetrad one-forms, which form a basis of the cotangent space, we then arrive at the expressions
\begin{subequations}\label{eq:varrels}
\begin{align}
\Sigma_a &= -\frac{1}{6}e_a{}^{\tau}\Theta_{\tau}{}^{\mu}\epsilon_{\mu\nu\rho\sigma}\dd x^{\nu} \wedge \dd x^{\rho} \wedge \dd x^{\sigma}\,,\\
\Delta_a &= \frac{1}{6}e_a{}^{\tau}\left(W_{\tau}{}^{\mu} - \nabla_{\omega}Y_{\tau}{}^{\mu\omega} + Y_{\tau}{}^{\mu\omega}M^{\omega}{}_{\psi\omega}\right)\epsilon_{\mu\nu\rho\sigma}\dd x^{\nu} \wedge \dd x^{\rho} \wedge \dd x^{\sigma}\,,\\
\Xi_a{}^b &= \frac{1}{6}e_a{}^{\tau}\theta^b{}_{\omega}Y_{\tau}{}^{\omega\mu}\epsilon_{\mu\nu\rho\sigma}\dd x^{\nu} \wedge \dd x^{\rho} \wedge \dd x^{\sigma}
\end{align}
\end{subequations}
for the three-forms we encountered in the tetrad formulation of teleparallel gravity theories. It is now straightforward to check that they indeed satisfy the Lorentz invariance conditions~\eqref{eq:lorinvmat} and~\eqref{eq:lorinvgrav}. By direct calculation we find
\begin{equation}\label{eq:connlorinv}
\begin{split}
\Sigma^{[a} \wedge \theta^{b]} &= -\frac{1}{6}\theta^{[a}{}_{\tau}\theta^{b]}{}_{\omega}\Theta^{\tau\mu}\epsilon_{\mu\nu\rho\sigma}\dd x^{\nu} \wedge \dd x^{\rho} \wedge \dd x^{\sigma} \wedge \dd x^{\omega}\\
&= -\theta^{[a}{}_{\tau}\theta^{b]}{}_{\omega}\Theta^{\tau\omega}\mathrm{vol}_{\theta}\\
&= -\theta^a{}_{\tau}\theta^b{}_{\omega}\Theta^{[\tau\omega]}\mathrm{vol}_{\theta}\\
&= 0
\end{split}
\end{equation}
and
\begin{equation}\label{eq:tetradlorinv}
\begin{split}
\Delta^{[a} \wedge \theta^{b]} - \eta^{c[a}\DD\Xi_c{}^{b]} &= \frac{1}{6}\bigg[\theta^{[a}{}_{\tau}\theta^{b]}{}_{\omega}g^{\tau\lambda}\left(W_{\lambda}{}^{\mu} - \nabla_{\psi}Y_{\lambda}{}^{\mu\psi} + Y_{\lambda}{}^{\mu\psi}M^{\phi}{}_{\psi\phi}\right)\epsilon_{\mu\nu\rho\sigma} + \partial_{\omega}\left(e_c{}^{\lambda}\eta^{c[a}\theta^{b]}{}_{\psi}Y_{\lambda}{}^{\psi\mu}\epsilon_{\mu\nu\rho\sigma}\right)\\
&\phantom{=}- e_d{}^{\lambda}\omega^d{}_{c\omega}\eta^{c[a}\theta^{b]}{}_{\psi}Y_{\lambda}{}^{\psi\mu}\epsilon_{\mu\nu\rho\sigma} + e_c{}^{\lambda}\eta^{c[a}\omega^{b]}{}_{d\omega}\theta^d{}_{\psi}Y_{\lambda}{}^{\psi\mu}\epsilon_{\mu\nu\rho\sigma}\bigg]\dd x^{\nu} \wedge \dd x^{\rho} \wedge \dd x^{\sigma} \wedge \dd x^{\omega}\\
&= \frac{1}{6}\bigg[\theta^{[a}{}_{\tau}\theta^{b]}{}_{\omega}g^{\tau\lambda}\left(W_{\lambda}{}^{\mu} - \nabla_{\psi}Y_{\lambda}{}^{\mu\psi} + Y_{\lambda}{}^{\mu\psi}M^{\phi}{}_{\psi\phi}\right)\\
&\phantom{=}+ e_c{}^{\lambda}\eta^{c[a}\theta^{b]}{}_{\psi}\left(\partial_{\omega}Y_{\lambda}{}^{\psi\mu} - \Gamma^{\phi}{}_{\lambda\omega}Y_{\phi}{}^{\psi\mu} + \Gamma^{\psi}{}_{\phi\omega}Y_{\lambda}{}^{\phi\mu} + \lc{\Gamma}^{\phi}{}_{\phi\omega}Y_{\lambda}{}^{\psi\mu}\right)\bigg]\epsilon_{\mu\nu\rho\sigma}\dd x^{\nu} \wedge \dd x^{\rho} \wedge \dd x^{\sigma} \wedge \dd x^{\omega}\\
&= \theta^{[a}{}_{\tau}\theta^{b]}{}_{\mu}g^{\tau\lambda}\left(W_{\lambda}{}^{\mu} - \nabla_{\psi}Y_{\lambda}{}^{\mu\psi} + Y_{\lambda}{}^{\mu\psi}M^{\phi}{}_{\psi\phi} + \partial_{\psi}Y_{\lambda}{}^{\mu\psi} - \Gamma^{\phi}{}_{\lambda\psi}Y_{\phi}{}^{\mu\psi} + \Gamma^{\mu}{}_{\phi\psi}Y_{\lambda}{}^{\phi\psi} + \lc{\Gamma}^{\phi}{}_{\psi\phi}Y_{\lambda}{}^{\mu\psi}\right)\mathrm{vol}_{\theta}\\
&= \theta^a{}_{\tau}\theta^b{}_{\mu}W^{[\tau\mu]}\mathrm{vol}_{\theta}\\
&= 0\,,
\end{split}
\end{equation}
using the symmetry of \(W^{\mu\nu}\) and \(\Theta^{\mu\nu}\). These relations will be instrumental when we show the equivalence of the field equations derived in the two different geometric frameworks.

Finally, we also translate the equations which we obtained from the assumption that each part of the action is invariant under diffeomorphisms. For the matter part~\eqref{eq:tetraddiffinvmat} then follows
\begin{equation}
0 = \lc{\DD}\Sigma_a = -\frac{1}{6}\left[\partial_{\omega}\left(e_a{}^{\tau}\Theta_{\tau}{}^{\mu}\epsilon_{\mu\nu\rho\sigma}\right) - \lc{\omega}^a{}_{b\omega}e_b{}^{\tau}\Theta_{\tau}{}^{\mu}\epsilon_{\mu\nu\rho\sigma}\right]\dd x^\omega \wedge \dd x^{\nu} \wedge \dd x^{\rho} \wedge \dd x^{\sigma} = -e_a{}^{\tau}\lc{\nabla}_{\mu}\Theta_{\tau}{}^{\mu}\mathrm{vol}_{\theta}\,,
\end{equation}
which is the energy-momentum conservation~\eqref{eq:metaffdiffinvmat}. Similarly, for the gravitational part one has
\begin{equation}
0 = \lc{\DD}\Delta_a + \DD\Xi_b{}^c \wedge (e_a \intprod M^b{}_c) = e_a{}^{\tau}\left[\lc{\nabla}_{\mu}\left(W_{\tau}{}^{\mu} - \nabla_{\omega}Y_{\tau}{}^{\mu\omega} + Y_{\tau}{}^{\mu\omega}M^{\omega}{}_{\psi\omega}\right) + M^{\nu}{}_{\mu\tau}\left(\nabla_{\psi}Y_{\nu}{}^{\mu\psi} - Y_{\nu}{}^{\mu\psi}M^{\phi}{}_{\psi\phi}\right)\right]\mathrm{vol}_{\theta}\,,
\end{equation}
which reproduces its metric-affine equivalent~\eqref{eq:metaffdiffinvgrav}.

\subsubsection{General teleparallel gravity}\label{sssec:relgen}
We now show that the field equations for the class of general teleparallel gravity theories derived in the metric-affine formulation in section~\ref{sssec:metaffgen} and in the tetrad formulation in section~\ref{sssec:tetradgen} agree. Using the relations~\eqref{eq:varrels} between the variation terms, this task becomes nearly trivial. In analogy to the calculation~\eqref{eq:tetradlorinv}, one finds that the connection field equation~\eqref{eq:spicongenfield} reduces to
\begin{equation}
0 = \DD\Xi_a{}^b = e_a{}^{\tau}\theta^b{}_{\mu}\left(\nabla_{\psi}Y_{\tau}{}^{\mu\psi} - Y_{\tau}{}^{\mu\psi}M^{\phi}{}_{\psi\phi}\right)\mathrm{vol}_{\theta}\,,
\end{equation}
and so reproduces the connection field equation~\eqref{eq:affcongenfield}. When this equation is satisfied, the tetrad field equation~\eqref{eq:tetradgenfield} simplifies and yields
\begin{equation}
0 = \Delta^a + \Sigma^a = \frac{1}{6}\theta^a{}_{\tau}\left(W^{\tau\mu} - \Theta^{\tau\mu}\right)\epsilon_{\mu\nu\rho\sigma}\dd x^{\nu} \wedge \dd x^{\rho} \wedge \dd x^{\sigma}\,,
\end{equation}
so that one obtains the metric field equation~\eqref{eq:metricgenfield}.

\subsubsection{Metric teleparallel gravity}\label{sssec:relmet}
We then proceed with the metric teleparallel class of theories. As argued in section~\ref{sssec:tetradmet}, the tetrad field equations take the same form~\eqref{eq:tetradgenfield} as in the general teleparallel case. Using only the relations~\eqref{eq:varrels}, without imposing any further equations to hold, one thus obtains
\begin{equation}
0 = \Delta^a + \Sigma^a = \frac{1}{6}\theta^a{}_{\tau}\left(W^{\tau\mu} - \nabla_{\omega}Y^{\tau\mu\omega} + Y^{\tau\mu\psi}T^{\omega}{}_{\omega\psi} - \Theta^{\tau\mu}\right)\epsilon_{\mu\nu\rho\sigma}\dd x^{\nu} \wedge \dd x^{\rho} \wedge \dd x^{\sigma}\,,
\end{equation}
where we used that the distortion tensor reduces to the contortion tensor in the absence of nonmetricity. We see that this result agrees with the combined metric and connection field equation~\eqref{eq:metaffmetfield}. Further, we have shown in section~\eqref{sssec:tetradmet} that the connection field equations~\eqref{eq:spiconmetfield} are identical to the antisymmetric part of the tetrad field equations. Indeed we see that the corresponding equation
\begin{equation}
0 = \DD\Xi^{[ab]} = -\theta^a{}_{\tau}\theta^b{}_{\mu}\left(\nabla_{\psi}Y^{[\tau\mu]\psi} - Y^{[\tau\mu]\psi}T^{\phi}{}_{\phi\psi}\right)\mathrm{vol}_{\theta}
\end{equation}
reproduces the antisymmetric part in the split~\eqref{eq:metaffmetfieldas}.

\subsubsection{Symmetric teleparallel gravity}\label{sssec:relsym}
Finally, we come to the case of symmetric teleparallel gravity. First, it is helpful to note that
\begin{equation}
0 = \DD\Xi_a{}^b = e_a{}^{\tau}\theta^b{}_{\mu}\left(\nabla_{\psi}Y_{\tau}{}^{\mu\psi} + \frac{1}{2}Y_{\tau}{}^{\mu\psi}Q_{\psi\phi}{}^{\phi}\right)\mathrm{vol}_{\theta}\,,
\end{equation}
and so
\begin{equation}
0 = e_b \intprod \DD\Xi_a{}^b = \frac{1}{6}e_a{}^{\tau}\left(\nabla_{\psi}Y_{\tau}{}^{\mu\psi} + \frac{1}{2}Y_{\tau}{}^{\mu\psi}Q_{\psi\phi}{}^{\phi}\right)\epsilon_{\mu\nu\rho\sigma}\dd x^{\nu} \wedge \dd x^{\rho} \wedge \dd x^{\sigma}\,.
\end{equation}
Hence, for the tetrad field equation~\eqref{eq:tetradsymfield} we find
\begin{equation}
0 = \Delta_a + \Sigma_a + e_b \intprod \DD\Xi_a{}^b = \frac{1}{6}e_a{}^{\tau}\left(W_{\tau}{}^{\mu} - \Theta_{\tau}{}^{\mu}\right)\epsilon_{\mu\nu\rho\sigma}\dd x^{\nu} \wedge \dd x^{\rho} \wedge \dd x^{\sigma}\,,
\end{equation}
which reproduces the metric field equation~\eqref{eq:metricgenfield}. For the connection field equation~\eqref{eq:spiconsymfield}, one has
\begin{equation}
0 = \DD(e_b \intprod \DD\Xi_a{}^b) = -e_a{}^{\tau}\left[\nabla_{\mu}\left(\nabla_{\psi}Y_{\tau}{}^{\mu\psi} + \frac{1}{2}Y_{\tau}{}^{\mu\psi}Q_{\psi\phi}{}^{\phi}\right) + \frac{1}{2}Q_{\mu\nu}{}^{\nu}\left(\nabla_{\psi}Y_{\tau}{}^{\mu\psi} + \frac{1}{2}Y_{\tau}{}^{\mu\psi}Q_{\psi\phi}{}^{\phi}\right)\right]\mathrm{vol}_{\theta}\,,
\end{equation}
which yields the connection field equation~\eqref{eq:affconsymfield}.


\section{Conclusion}\label{sec:conclusion}
We have derived the field equations for the three generic families of teleparallel gravity theories, in which one of the dynamical field variables is a connection with vanishing curvature, but non-vanishing torsion or nonmetricity, or both. For each family, we have used two different geometric formulations, either in terms of a metric and affine connection, or through a tetrad and spin connection. Further, we have used different methods to implement the constraints on the connection variable in the derivation of the field equations, either by using Lagrange multipliers in the action, or by restricting the variation of the field variables such that they preserve the constraint. By comparing the results, we have shown explicitly the equivalence of the different geometric formulations and variation prescriptions for each of the three classes of teleparallel gravity theories.

For the tetrad formulation, we have chosen the Lorentz covariant formulation of teleparallel gravity theories~\cite{Krssak:2018ywd}, in which the action is invariant under combined Lorentz transformation of the tetrad and the spin connection. Further, we have assumed that the action functional is invariant under diffeomorphisms. From this assumption we have derived a number of geometric identities which are automatically satisfied by the terms obtained by varying the gravitational part of the action. In particular, we have shown that in the case of symmetric teleparallel gravity theories, these identities are fully sufficient to obtain the connection field equations from the metric ones. This finding supports the interpretation of the connection as a gauge degree of freedom related to diffeomorphism invariance in these theories.

While we have focused on the derivation of the gravitational field equations by variation of teleparallel gravity actions, and shown the equivalence of different approaches for this task, one may wonder whether this equivalence also holds beyond the level of the classical field equations. Possible applications include the Casimir effect~\cite{Ulhoa:2019roa,Bahamonde:2019yyp} and the entropy of black holes~\cite{Ulhoa:2019roa,Ulhoa:2021ory}, where also surface terms in the action are relevant. Further, teleparallel gravity theories may also be studied within more general geometric frameworks, such as higher gauge theory~\cite{Baez:2012bn} or Cartan geometry~\cite{Hohmann:2015pva,Fontanini:2018krt,LeDelliou:2019esi,Huguet:2020ler,Huguet:2021roy}. Another possible extension is to study teleparallel gravity theories which are not defined by an action functional, but in the premetric approach via a constitutive relation~\cite{Hehl:2016glb,Itin:2016nxk,Hohmann:2017duq,Itin:2018dru,Koivisto:2021jdy}.

In this work we have considered teleparallel gravity theories within the realm of classical field theory, and so one may wonder whether our results can also be extended to the quantum level. The latter is not necessarily the case, since although we have demonstrated that various geometric formulations yield the same classical field equations and space of solutions, they feature different field variables, from which quantum corrections may arise. This question is of particular interest for teleparallel gravity theories which reproduce general relativity at the classical level~\cite{BeltranJimenez:2019tjy,Jimenez:2019ghw,Bohmer:2021eoo}. While one may expect the same difficulties to arise which obstruct a quantization of general relativity in its curvature formulation, teleparallel gravity theories may offer a better approach to solve these difficulties through a suitable quantum modified gravity theory.

\vspace{6pt}

\funding{This work was supported by the Estonian Research Council through the Personal Research Funding project PRG356 and by the European Regional Development Fund through the Center of Excellence TK133 ``The Dark Side of the Universe''.}

\acknowledgments{The author thanks Sebastian Bahamonde and Jackson Levi Said for the kind invitation to contribute to this Special Issue.}

\conflictsofinterest{The author declares no conflict of interest.}

\reftitle{References}


\externalbibliography{yes}
\bibliography{televar}
\end{document}